\documentclass[11pt]{article}
\usepackage[T1]{fontenc}
\usepackage[utf8]{inputenc}
\usepackage{ae,lmodern}
\usepackage{array}
\usepackage{longtable}
\usepackage{float}
\usepackage{booktabs}
\usepackage{units}
\usepackage{bm}
\usepackage{multirow}
\usepackage{lscape}
\usepackage{amsmath}
\usepackage{amssymb}
\usepackage{graphicx}
\usepackage{setspace}
\usepackage[authoryear]{natbib}
\usepackage{ccaption}
\usepackage{gensymb}

\makeatletter

\providecommand{\tabularnewline}{\\}
\usepackage{amsthm}
\usepackage{bm}
\usepackage{fancyhdr}
\usepackage{nicefrac}
\usepackage{setspace}

\usepackage{fullpage}
\usepackage{lineno}
\usepackage{caption}
\usepackage{authblk}
\usepackage{makecell}
\usepackage{rotating}
\usepackage{soul}
\usepackage{longtable}
\usepackage{lineno}

\usepackage{appendix}
\usepackage{hyperref}
\usepackage{authblk}
\usepackage{xcolor}

\hypersetup{
    colorlinks=true,
    citecolor=purple,
    linkcolor=blue,
    filecolor=magenta,      
    urlcolor=cyan,
}

\makeatother

\begin{document}

\title{The effect of seasonal strength and abruptness on predator--prey dynamics}

\author[1]{Alix M. C. Sauve\thanks{Corresponding author: alix.sauve@cri-paris.org}}
\author[2]{Rachel A. Taylor}
\author[1,3]{Frederic Barraquand}

\affil[1]{University of Bordeaux, Integrative and Theoretical Ecology, LabEx COTE, France}
\affil[2]{Animal and Plant Health Agency (APHA), Weybridge, United Kingdom}
\affil[3]{CNRS, Institute of Mathematics of Bordeaux, France}

\date{}

\maketitle

\vfill
This manuscript is published in \textit{Journal of Theoretical Biology}. Cite as:

Alix M.C. Sauve, Rachel A. Taylor, Frédéric Barraquand,
The effect of seasonal strength and abruptness on predator–prey dynamics, Journal of Theoretical Biology, Volume 491, 2020, 110175.

\href{https://doi.org/10.1016/j.jtbi.2020.110175}{https://doi.org/10.1016/j.jtbi.2020.110175}

\pagebreak

\begin{abstract}
Coupled dynamical systems in ecology are known to respond to the seasonal forcing of their parameters with multiple dynamical behaviours, ranging from seasonal cycles to chaos. Seasonal forcing is predominantly modelled as a sine wave. However, the transition between seasons is often more sudden as illustrated by the effect of snow cover on predation success. A handful of studies have mentioned the robustness of their results to the shape of the forcing signal but did not report any detailed analyses. Therefore, whether and how the shape of seasonal forcing could affect the dynamics of coupled dynamical systems remains unclear, while abrupt seasonal transitions are widespread in ecological systems. To provide some answers, we conduct a numerical analysis of the dynamical response of predator--prey communities to the shape of the forcing signal by exploring the joint effect of two features of seasonal forcing: the magnitude of the signal, which is classically the only one studied, and the shape of the signal, abrupt or sinusoidal. We consider both linear and saturating functional responses, and focus on seasonal forcing of the predator's discovery rate, which fluctuates with changing environmental conditions and prey's ability to escape predation. Our numerical results highlight that a more abrupt seasonal forcing mostly alters the magnitude of population fluctuations and triggers period--doubling bifurcations, as well as the emergence of chaos, at lower forcing strength than for sine waves. Controlling the variance of the forcing signal mitigates this trend but does not fully suppress it, which suggests that the variance is not the only feature of the shape of seasonal forcing that acts on community dynamics. Although theoretical studies may predict correctly the sequence of bifurcations using sine waves as a representation of seasonality, there is a rationale for applied studies to implement as realistic seasonal forcing as possible to make precise predictions of community dynamics.
\end{abstract}


\pagebreak

\section{Introduction}

Seasonality has long been thought to be a key driver of population cyclicity, both annual and multi--annual \citep{bjornstad_geographic_1995,taylor_how_2013,barraquand_moving_2017}, and its
dynamical implications have been the subject of several theoretical studies \citep[e.g.,][]{kuznetsov_bifurcations_1992,rinaldi_multiple_1993,gragnani_universal_1995,king_geometry_2001}. Most implementations of seasonality consist of a periodic forcing on parameters driving reproduction, carrying capacity, mortality, or attack rates \citep{rinaldi_multiple_1993,taylor_how_2013}, which are likely responding to environmental fluctuations. Many models produce cyclic dynamics which can develop into erratic or unexpectedly large fluctuations, i.e., chaos and resonances \citep[e.g.,][]{rinaldi_multiple_1993,sabin_chaos_1993,king_weakly_1996}. This is true even for two--species predator--prey systems, that are known for the most part to exhibit either a fixed point equilibrium or a limit cycle when having autonomous dynamics. Seasonality therefore has a major effect on the dynamics of predator--prey systems modelled as nonlinear coupled differential equations. 

\begin{figure}[H]
\centering \includegraphics[scale=0.8]{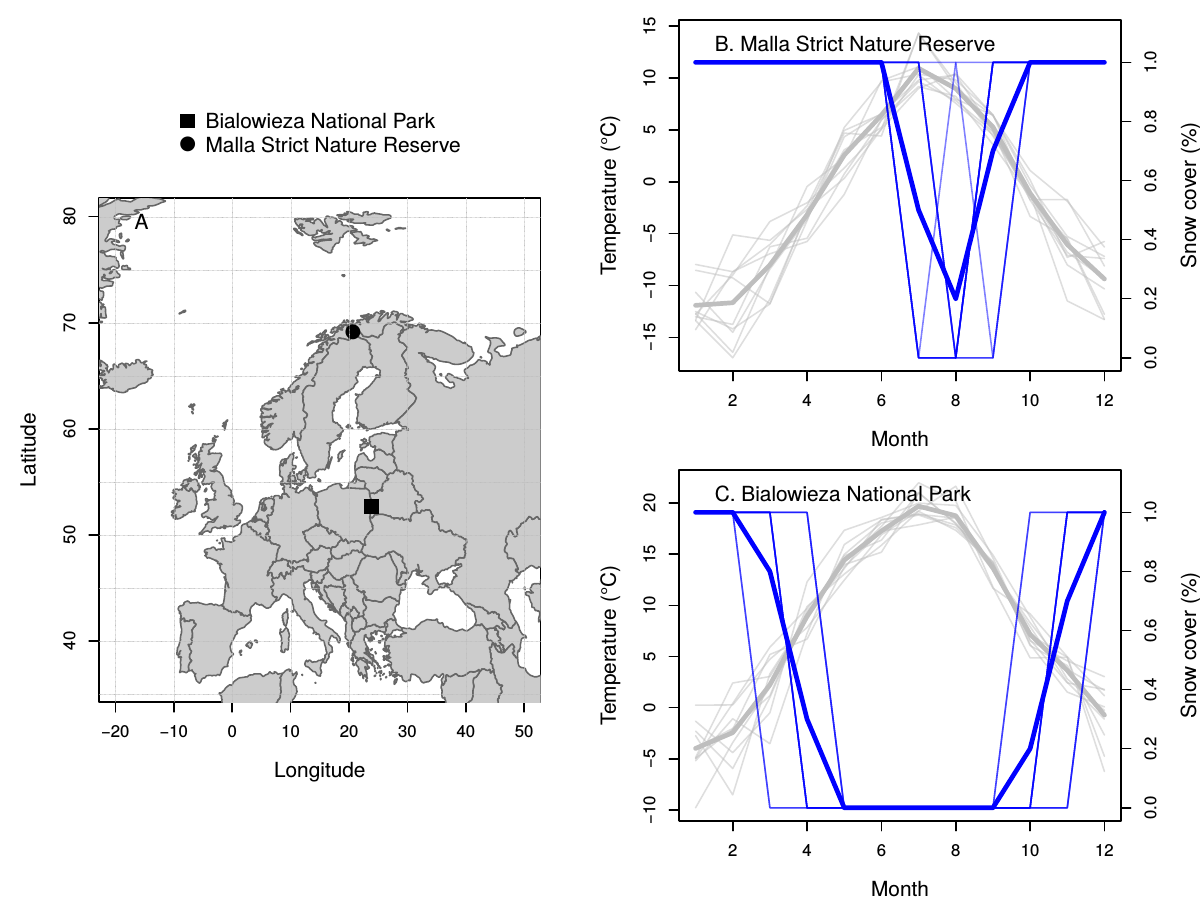}
\caption{Environmental forcing may be smooth or abrupt depending on the climatic variable affecting 
ecological process rates. Temperature (grey lines in B--C) usually follows a wave-like pattern while 
snow cover (blue lines in B--C) produces an abrupt environmental forcing. Climatic variables are 
represented at two different locations varying in latitude (A): B) Malla Strict Nature Reserve 
($69^{\degree}3'56''\textrm{N}\; 20^{\degree}40'16''\textrm{E}$) and C) Bia{\l}owie{\.z}a National 
Park ($52^{\degree}45'7.66''\textrm{N}\;23^{\degree}52'44.86''\textrm{E}$).
B and C contain modified Copernicus Climate Change Service Information \citep{era5}.
Each thin line corresponds to monthly--averaged temperature or snow cover for a given year 
(2009--2018), and thick lines to their average value across years. Months are indexed from January (1) 
to December (12). Snow cover is expressed as the percentage area covered with snow within the grid box 
matching locations and is based on snow depth estimated in meters of water equivalent \citep{ecmwf_part_2016}.}
\label{fig:ClimateForcing}
\end{figure}

For studies aiming at mapping dynamical behaviours onto the parameter space thanks to bifurcation analysis, focusing on a sinusoidal forcing is technically convenient. This type of forcing preserves the continuity of functions describing species growth rates, and enables numerical continuation for bifurcation analyses by writing the sine wave forcing as a built--in harmonic oscillator \citep[e.g.,][]{doedel_auto:_1981,dhooge_matcont:_2003}. An unfortunate side effect of this technical choice is that the bulk of the theoretical literature pertains to the dynamical consequences of the magnitude of seasonal forcing while assuming a match with the seemingly wave--like pattern of annual temperatures (Fig. \ref{fig:ClimateForcing}).

Although temperature is a prominent feature of climate seasonality, there is 
no reason to believe that the temporal changes of ecological parameters 
always reproduce exactly those of temperature. Snowfalls and extent of snow cover typically display more abrupt temporal patterns as they are not 
governed purely by temperature \citep[Fig. \ref{fig:ClimateForcing},][]{mott_seasonal_2018}. Depending on its features (depth, density), snow cover 
can affect access to food sources \citep{korslund_small_2006} and to prey 
\citep{bilodeau_effect_2013,fauteux_seasonal_2015}, especially for generalist 
predators \citep{hansson_gradients_1985,sonerud_effect_1986}, but also 
hinder species movements and killing success 
\citep{stenseth_snow_2004,penczykowski_winter_2017}. Foraging and 
reproduction rates within ecological communities living in snowy landscapes 
are thus likely to undergo abrupt seasonal forcings. Other abrupt forcings 
on populations may occur seasonally. For instance, population changes due to 
migration or dormancy can be abrupt as they are usually triggered by 
temperature thresholds and resource availability 
\citep{saino_climate_2010,pau_predicting_2011,therrien_irruptive_2014}. Prey 
behaviour such as investing in reproduction can, from one day to the next, 
increase the prey vulnerability to predators by making them suddenly more 
conspicuous \citep{magnhagen_predation_1991}. Predators can also 
specifically target the breeding period regardless of prey abundance because 
juveniles are easier to hunt \citep{tornberg_prey_1997,probst_aerial_2011}.

That environmental forcing has some level of abruptness is thus well--established empirically. In fact, some  empirically--driven rodent modelling studies (rodents--mustelids or plants--rodents) do include abrupt seasonality already \citep[e.g.,][]{hanski_population_1993,gilg_cyclic_2003,reynolds_comparison_2013}. The predator functional response may also shift abruptly, to mimic the ability of some predators to switch between specialist and generalist foraging patterns from one season to another \citep{tyson_seasonally_2016}.
These ``abrupt seasonality'' models are able to produce a similar range of dynamical behaviours to their ``sinusoidal seasonality'' counterparts. But bifurcation studies tend to neglect more abrupt signals for the technical reasons we mentioned above. Only a few studies mention the robustness of their results to the shape of seasonal shifts in parameter values, but without providing any actual proof of such robustness (e.g., \citealt{turchin_empirically_1997}, for predator--prey models; \citealt{greenman_external_2004,ireland_effect_2004}, for epidemiological models).
Yet, as there are empirical evidences of abrupt species response to environmental seasonality, the shape of the forcing signal cannot be dismissed as a mere modelling detail without further investigation of its contribution to population dynamics.

In the present study, we address this gap by exploring the effect of the abruptness of the seasonal transition on the dynamical behaviour of a predator--prey community.
Inspired by potential effects of snow properties on predation rates, we implement a more or less abrupt seasonal forcing of the predator discovery rate.
We start with a reminder on unforced predator--prey dynamics when governed by a type I and a type II functional response, followed by a brief study on their response to a sinusoidal forcing on discovery rates mimicking seasonal variation in predator's ability to successfully hunt prey.
Then, we explore the joint effect of the magnitude (how much a seasonal parameter varies) and the shape (from sinusoidal to rectangular) of seasonal forcing on the predator's discovery rate, and test whether the differences between forcing shapes that arise are explained by the signal variance.

\section{Modelling predator--prey communities in a seasonal world}

\subsection{Unforced predator--prey models}

\begin{figure}[H]
\centering \includegraphics[scale=0.6]{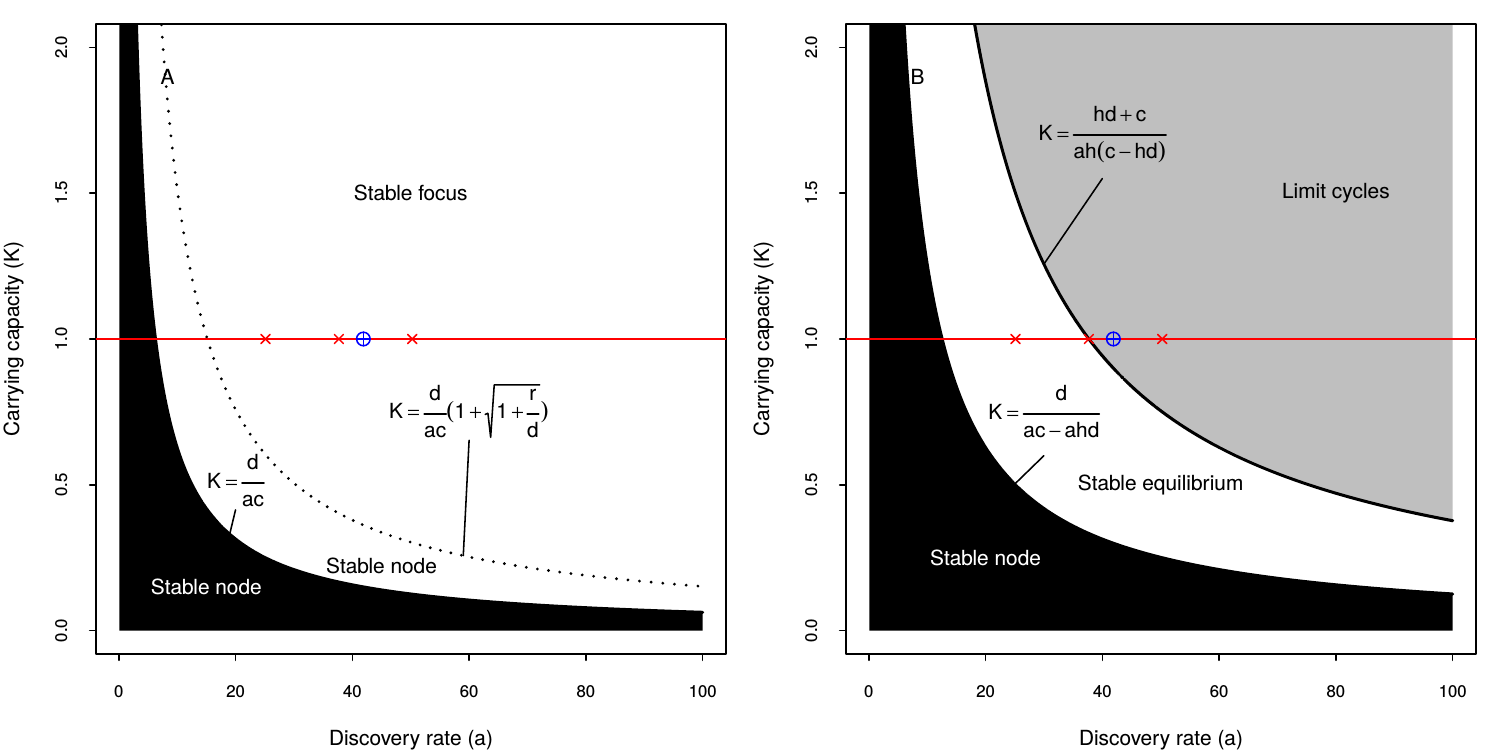}
\caption{Equilibrium state of eq. \eqref{eq:system} with respect to the carrying capacity $K$ and discovery rate $a$. Predation is either ruled by A) a type I or B) a type II functional response. The white area delineates conditions of stable coexistence, the black area corresponds to conditions under which only the prey survives, and the grey area those under which species coexist in limit cycles. Limits of these areas are drawn for the parameterisation described in Table \ref{tab:ParamValues}. The parameter sets we explore in the present study are indicated with red crosses along $K=1$. Parameterisation from \citet{rinaldi_multiple_1993} is marked with a blue crossed circle.}
\label{fig:EquilibriumState}
\end{figure}

We describe the dynamics of the prey density $x$ and the predator density $y$ with a Gause--type model \citep{kuang_nonuniqueness_1988}:

\begin{equation}
\left\{ \begin{array}{l}
\frac{dx}{dt}=r\left(1-\frac{x}{K}\right)x-f(x)y\\
\frac{dy}{dt}=cf(x)y-dy
\end{array}\right.\label{eq:system}
\end{equation}

\noindent where $r$ is the intrinsic growth rate of the prey; $K$, prey carrying capacity; $c$, the conversion efficiency of the predator; $d$, the mortality rate of the predator; and $f(x)$ is the functional response of the predator to prey density. Hence, the prey density grows logistically in the absence of the predator. Herein, we consider both type I and type II functional responses to test whether density--dependent predation alters the community response to the shape of the signal. A functional response of type I assumes that the intake of prey is proportional to prey density only, and so a linear term is used ($f(x)=ax$, $a$ being the discovery rate). This model allows the coexistence of both populations, provided that $K$ is large enough ($K > \frac{d}{ac}$), with dynamics converging to a stable node or a stable focus depending on parameter values (Fig. \ref{fig:EquilibriumState}A, and \ref{sec:UnforcedSystem}). A functional response of type II describes an intake of prey which increases linearly at low prey density and saturates at higher densities:

\begin{equation}
f(x)=\frac{ax}{1+ahx}\label{eq:TypeII_Holling}
\end{equation}

\noindent where $h$ is the predator's handling time and $a$ is once again the discovery rate. The type II functional response allows for more diverse dynamics as the community may converge to a stable equilibrium, either by monotonic or oscillatory decay, or to a limit cycle \citep[Fig. \ref{fig:EquilibriumState}B and \ref{sec:UnforcedSystem};][]{rosenzweig_graphical_1963}.

\subsection{Implementation of a seasonal forcing on attack rates}

For both models, we assume the discovery rate $a$ to change over time to represent a fluctuating environment which alternatively facilitates or impedes predation by acting on individuals' mobility and ability to hide \citep{kuznetsov_bifurcations_1992}. This partly corresponds to \citet{rinaldi_multiple_1993}'s fifth seasonality mechanism by which the half saturation constant $b=\nicefrac{1}{ah}$ periodically fluctuates: depending on the season, the intake of prey increases more or less steeply with prey density.

We force the discovery rate with a time--dependent function

\begin{equation}
a(t)=\bar{a}(1+s(t))
\label{eq:SeasonalA}
\end{equation}

\noindent where $\bar{a}$ is the mean discovery rate, and $s(t)$ is a periodic function with period $\mathcal{P}=1\mathit{~year}$ and such that $|s(t)|\leq1$. We refer to seasonality with two of its features in mind: its amplitude \citep[e.g.,][]{taylor_seasonal_2013}, and the type of transition between summer and winter \citep[e.g.,][]{turchin_empirically_1997}. The first tells us how different the two seasons are regarding the parameters affected by seasonality, while the second indicates how abrupt the seasonal shift is. To account for both of these aspects of seasonality, we write

\begin{equation}
s(t, \epsilon, \theta)=\frac{\epsilon\sin(2\pi t)}{|\sin(2\pi t)|}|\sin(2\pi t)|^{\theta}
\label{eq:period_func}
\end{equation}

\noindent where $\epsilon$ ($\epsilon\:\in\left[0,1\right]$) is a dimensionless parameter capturing the forcing strength, i.e. the amplitude of seasonal forcing, and $\theta\:\left(\theta\in\left[0,1\right]\right)$ describes the abruptness of the seasonal transition. Whilst $\theta$ values close to $1$ correspond to a simple sine wave producing a smooth transition between summer and winter (following the temporal pattern of temperatures as in Fig. \ref{fig:ClimateForcing}; e.g. \citealp{rinaldi_multiple_1993,turchin_empirically_1997}), $\theta$ close to $0$ corresponds to a step function depicting very abrupt seasonal shifts (e.g., changes in snow cover, Fig. \ref{fig:ClimateForcing}). To investigate the effect of seasonality on community stability and persistence, we describe the community dynamics over $\epsilon\in\left[0,1\right]$ and $\theta\in\left[0,1\right]$. Hereafter, we refer to the season during which predation is maximal as ``the predation season'' (i.e., when $s(t, \epsilon, \theta) > 0$). This can correspond to the period of time during which snow cover is minimal, or prey reproduction is higher.

\subsection{Simulation setup}

Parameter values are identical to those used by \citet{rinaldi_multiple_1993} and later by \citet{taylor_seasonal_2013}, and are detailed in Table \ref{tab:ParamValues}.

\begin{table}[H]
\begin{centering}
{\small{}}
\global\long\def\arraystretch{1.5}%
{\small{} \setlength{\tabcolsep}{3pt} }%
\begin{tabular}{cccc}
\hline 
\textbf{\small{}Parameter}{\small{} } & \textbf{\small{}Meaning}{\small{} } & \textbf{\small{}Unit}{\small{} } & \textbf{\small{}Value}\tabularnewline
\hline 
\textit{\small{}x, y}{\small{} } & {\small{}Species density } & {\small{}$N.km^{-2}$ } & {\small{}- }\tabularnewline
\hline 
\textit{\small{}r}{\small{} } & {\small{}Intrinsic growth rate of the prey } & {\small{}$y^{-1}$ } & {\small{}$2\pi$ }\tabularnewline
\hline 
\textit{\small{}K}{\small{} } & {\small{}Carrying capacity of the prey } & {\small{}$N.km^{-2}$ } & {\small{}$1$ }\tabularnewline
\hline 
\textit{\small{}a}{\small{} } & {\small{}Discovery rate } & {\small{}$N^{-1}.km^{2}.y^{-1}$ } & {\small{}- }\tabularnewline
\hline 
\textit{\small{}h}{\small{} } & {\small{}Handling time} & {\small{}$y$ } & {\small{}$\nicefrac{1}{4\pi}^{*}$ }\tabularnewline
\hline 
\textit{\small{}c}{\small{} } & {\small{}Coefficient of conversion } & {\small{}- } & {\small{}$1$ }\tabularnewline
\hline 
\textit{\small{}d}{\small{} } & {\small{}Mortality rate of the predator } & {\small{}$y^{-1}$ } & {\small{}$2\pi$ }\tabularnewline
\hline 
\end{tabular}{\small\par}
\par\end{centering}
{\small{}\caption{\label{tab:ParamValues}Parameterisation of the community models.\protect \\
{*}The handling time $h$ is parameterised as the inverse of the
reference value for the maximum killing rate in \citet{rinaldi_multiple_1993}.}
}{\small\par}
\end{table}

To explore the effect of seasonal forcing on the predator's discovery rate, we simulate the dynamics of the predator--prey community with varying values of $\epsilon\in\left[0,1\right]$ with increment $0.01$, and $\theta\in\left[0,1\right]$ with increment $0.1$. As the forcing signal is exerted on the discovery rate $a$ which is a bifurcation parameter, we consider different values of $\bar{a}$ that produce different dynamical behaviours in the unforced system (Fig. \ref{fig:EquilibriumState} and Fig. \ref{fig:BifurcDiagOna}).

The time unit is one year, and dynamics are simulated over at least $2000$ years so that they reach asymptotic dynamics.
We use Matlab's ODE solver \textit{ode15s} to integrate equations, with relative error tolerance set at $\mathit{Reltol}=10^{-8}$ and absolute error tolerance at $\mathit{Abstol}=10^{-6}$, and record one time step every week ($t_{step}=\nicefrac{1}{52}\;year$) to examine time series.
We also use the option \textit{`NonNegative'} to constrain the system to interior solutions. Moving beyond a sinusoidal forcing as described in eq. \eqref{eq:period_func} introduces discontinuities in system's eq. \eqref{eq:system} when $t=0\;(mod\;\frac{1}{2})$. To avoid building errors on these discontinuities, we interrupt the numerical integration at the end of each season (i.e., at $t=0\;(mod\;\frac{1}{2})$).
The form of the forcing function $s(t)$ also prevents bifurcation analysis by means of numerical continuation because it cannot be recast into a simple harmonic oscillator to transform the system into an autonomous one \citep[as in][]{doedel_auto:_1981,dhooge_matcont:_2003}.
This constrains our study to numerical integration of seasonally forced ODEs and graphical analysis of the attractors. In addition to phase portraits, we inspect the attractors by sampling the time series annually and by displaying them with numerical bifurcation diagrams (similar to those of discrete--time systems) and stroboscopic maps (projecting the attractor onto a Poincaré section). 

\section{The dynamical response to the magnitude of the seasonal forcing $\epsilon$\label{sec:Epsilon}}

Although mostly focused on prey reproduction rather than predation terms, the dynamical consequences of the magnitude of a sinusoidal seasonal forcing has been explored in detail in previous studies \citep[e.g.,][]{kuznetsov_bifurcations_1992,rinaldi_multiple_1993,king_weakly_1996,taylor_seasonal_2013}.
Consequently, this section serves more as a reminder of the expected dynamics and bifurcations under sinusoidal seasonal forcing to support the interpretation of results in the following section.

\begin{figure}[H]
\centering
\includegraphics[scale=0.75, trim = {2cm 0 2cm 0}, clip]{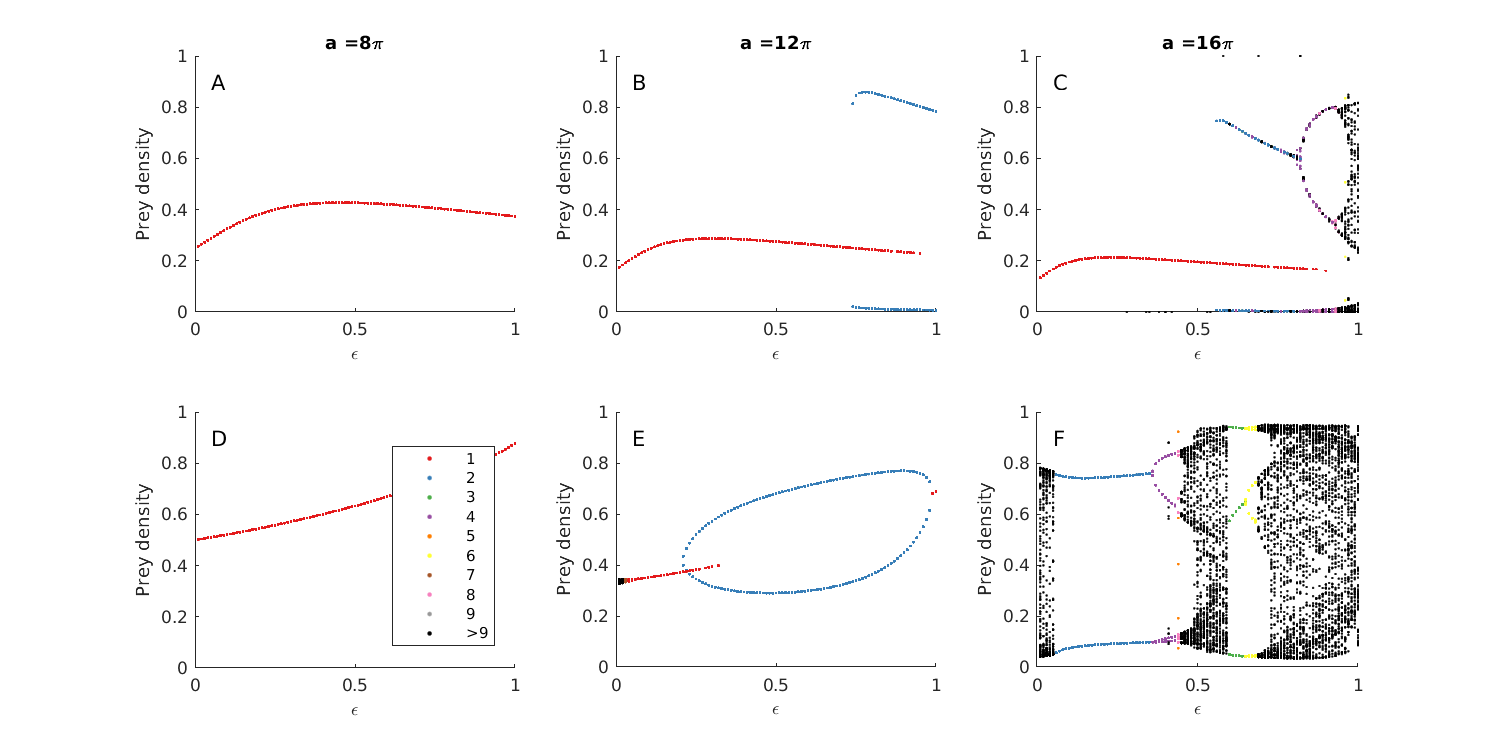}
\caption{Prey density at the beginning of each predation season ($s(t, \epsilon, 1) > 0$, 10 years recorded) against the magnitude of the seasonal forcing $\epsilon$, with predation governed by A--B--C) a type I functional response, and D--E--F) a type II functional response.
Each plot corresponds to a value of the mean discovery rates $\bar{a}$: $8\pi$ (A and D), $12\pi$ (B and E), and $16\pi$ (C and F).
Each point corresponds to a given simulation with a specific set of initial conditions.
For each value of $\epsilon$, we ran 10 simulations initiated with different densities ($(x_0,\,y_0) \in ]0,\,1]$).
Point colour corresponds to the system periodicity estimated with \citet{taylor_how_2013}'s algorithm, with black points indicating periodicity over 9 years, quasi--periodic or chaotic regimes.}
\label{fig:BifurcDiagOnepsilon_Prey}
\end{figure}

Sinusoidal forcing (i.e., $\theta = 1$) of the predator's discovery rate is the source of fluctuating dynamics with multiple periods, for both type I or type II functional responses (Fig. \ref{fig:BifurcDiagOnepsilon_Prey} and Fig. \ref{fig:BifurcDiagOnepsilon_Pred} in \ref{sec:TS_BifurcDiagEpsilon}). In our simulation setup, community dynamics depend both on the mean discovery rate $\bar{a}$ that determines the dynamical behaviour of the unforced system (Fig. \ref{fig:EquilibriumState}), and the magnitude $\epsilon$ of the seasonal forcing exerted on this parameter, which we detail below.

When predation is ruled by a type I functional response, the unforced system is always a stable point. As a result, seasonal predation mostly produces annual cycles, at least when the forcing strength $\epsilon$ is low (Fig. \ref{fig:BifurcDiagOnepsilon_Prey}A to C and Fig. \ref{fig:BifurcDiagOnepsilon_Pred}). However, higher mean discovery rates $\bar{a}$ allow more complex dynamics (Fig. \ref{fig:BifurcDiagOnepsilon_Prey}B and C). At stronger seasonal forcing, one year and two year cycles coexist (see Fig. \ref{fig:Bassins} for the illustrations of basins of attractions and Fig. \ref{fig:TS_EpsilonOnly} for the corresponding dynamics). If $\bar{a}$ is large enough, successive period--doubling bifurcations, which produce cycles of period twice larger than previous cycles, lead to chaotic behaviour (Fig. \ref{fig:BifurcDiagOnepsilon_Prey}C).

When predation is ruled by a type II functional response, the forced system can exhibit even more complex dynamics (Fig. \ref{fig:BifurcDiagOnepsilon_Prey}D to F). Again, if the unforced dynamics are a stable point, seasonal forcing on the predator's discovery rate generates annual cycles (Fig. \ref{fig:BifurcDiagOnepsilon_Prey}D). On the other hand, when the unforced system is a limit cycle (Fig. \ref{fig:EquilibriumState}B), seasonal predation leads, for low values of the forcing strength, to quasi--periodic behaviour (Fig. \ref{fig:BifurcDiagOnepsilon_Prey}E and F). A larger seasonal forcing $\epsilon$ then produces a stable cycle with a period similar to the unforced orbit (Fig. \ref{fig:BifurcDiagOnepsilon_Prey}E and F). When the functional response is of type II, seasonally--forced predation allows the coexistence of orbits with different periods as well (e.g., Fig. \ref{fig:Bassins} and \ref{fig:TS_EpsilonOnly}), sometimes with periods higher than one and two years (Fig. \ref{fig:ZoomBifurcDiagEpsilon_TypeII_a16pi_E0305}). As we further increase the forcing strength $\epsilon$ when the discovery rate is high, community dynamics become chaotic through successive period--doubling bifurcations (Fig. \ref{fig:BifurcDiagOnepsilon_Prey}F, Fig. \ref{fig:ZoomBifurcDiagEpsilon_TypeII_a16pi_E0305} and \ref{fig:ZoomBifurcDiagEpsilon_TypeII_a16pi_E065075}). In addition, we observe a large periodic window in between chaotic regimes when predation has a saturating functional response (Fig. \ref{fig:BifurcDiagOnepsilon_Prey}F). In this periodic window, the cycling dynamics have a period of three years and then of six years.

We have thus numerically verified the dynamical behaviour of seasonally forced predator--prey communities that have been shown to occur in models with either type I or type II responses \citep[see][]{rinaldi_multiple_1993,king_weakly_1996}. Indeed, \cite{king_weakly_1996} highlight the coexistence of multiple attractors for dissipative predator--prey systems with a type I functional response (Lotka--Volterra model with self--regulation in the prey), while \cite{vandermeer_seasonal_1996} found a Feigenbaum cascade (or period--doubling route) to chaos. For the models with type II functional response, the ``universal'' bifurcation diagram of \cite{rinaldi_multiple_1993} and \cite{kuznetsov_bifurcations_1992}'s works shed light on the various dynamical behaviours. At low forcing strengths in the vicinity of the Hopf bifurcation, a fold bifurcation gives birth to a pair of two--year cycles, one stable and the other unstable. This explains the sudden appearance of a stable two--year cycle coexisting with the annual cycle in Fig. \ref{fig:BifurcDiagOnepsilon_Prey}E. When further increasing the forcing strength, coexistence of multiple cycles stops: the system crosses a (subcritical) period--doubling bifurcation by which the stable cycle of period one year collides with the unstable cycle of period two to become unstable \citep{rinaldi_multiple_1993,dercole_dynamical_2011}. This gives way to the stable two--year cycle which previously emerged at the fold bifurcation \citep[Fig. \ref{fig:BifurcDiagOnepsilon_Prey}E, ][]{rinaldi_multiple_1993}. For higher discovery rates, we observe a region of chaotic behaviour which appears through multiple period--doubling bifurcations. The three--year cycles in between two chaotic regions which we observe in Fig. \ref{fig:BifurcDiagOnepsilon_Prey}F are the result of a fold bifurcation being crossed while increasing $\epsilon$. The system then reverts back to chaos through multiple period--doubling bifurcations.

\section{Dynamical response to the shape of seasonal forcing}
\label{sec:EpsilonTheta}

\begin{figure}[H]
\centering \includegraphics[scale=0.75, trim = {1.5cm 1cm 2cm 0}, clip]{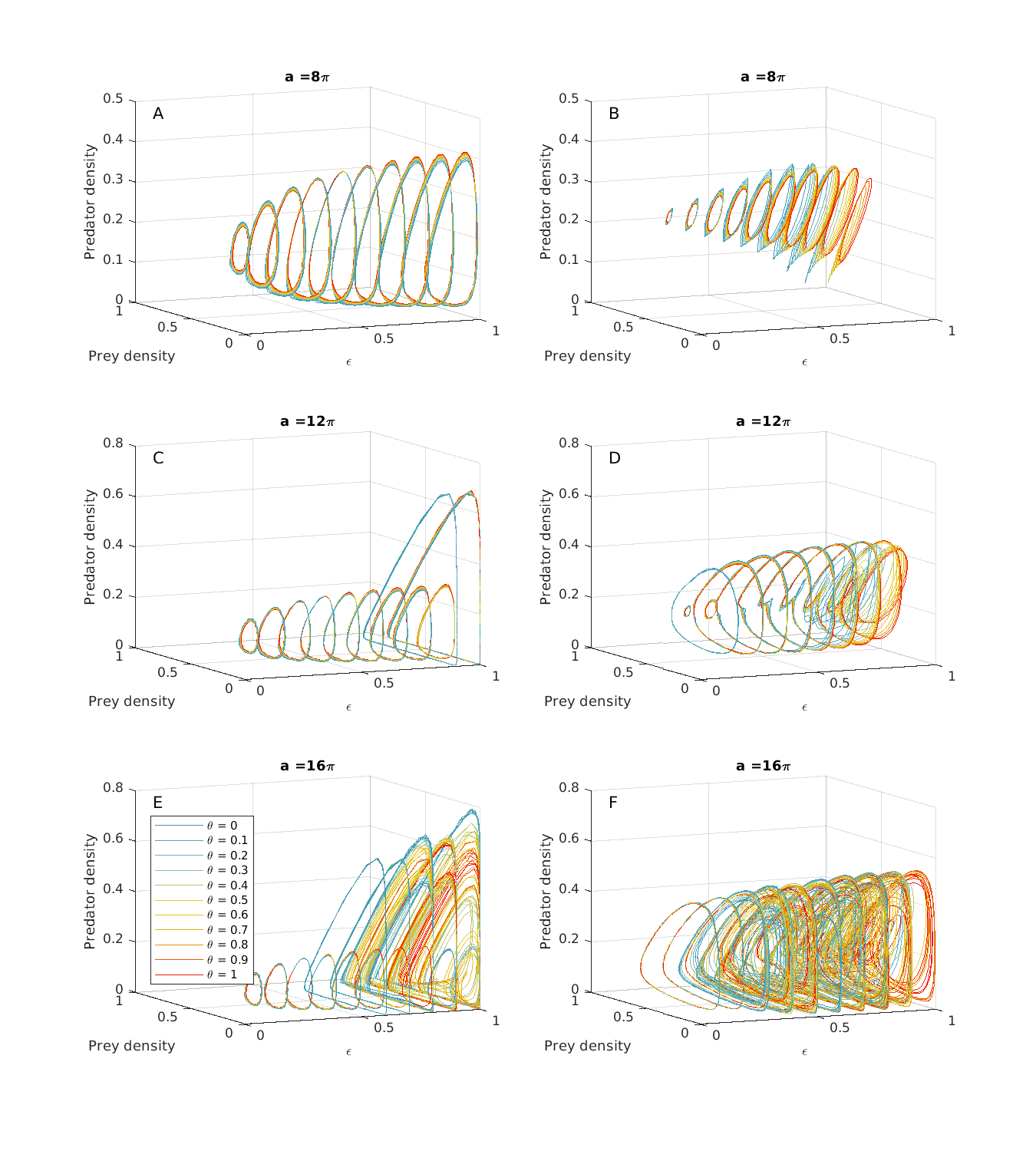}
\caption{Phase portraits of the predator--prey community dynamics undergoing seasonal forcing of various magnitudes $\epsilon$, and abruptness $\theta$ ($\theta\rightarrow0$: more abrupt forcing, $\theta\rightarrow1$: smoother forcing).
Predation is either shaped by a type I functional response (left column: A, C and E), or by a type II functional response (right column: B, D and F).
We consider different mean discovery rates: A--B) $\bar{a}=8\pi$, C--D) $\bar{a}=12\pi$, and E--F) $\bar{a}=16\pi$.
Simulations are all initiated with the same densities $(x_0,y_0) = (0.3,0.3)$.
 Only asymptotic dynamics covering the last ten years of simulation are displayed. Closed curves correspond to periodic dynamics while other attractors are indicative of quasi--periodic and chaotic behaviours.}
\label{fig:PP_OnThetaEpsilon} 
\end{figure}

Changing the shape of the seasonal forcing, $\theta$, can alter the predator--prey trajectory despite identical parameterisation and initial conditions (Fig. \ref{fig:PP_OnThetaEpsilon}), and these alterations are of different nature depending on the values of the $\{\bar{a},\,\epsilon,\,\theta\}$--tuple.

When the mean discovery rate $\bar{a}$ is such that the system does not cross bifurcations when changing the value of $\epsilon$ ($\bar{a} = 8\pi$, Fig. \ref{fig:BifurcDiagOnepsilon_Prey}A and D), a more abrupt seasonal forcing on the predator's discovery rate ($\theta\rightarrow0$) hardly deviates the community trajectory (e.g., Fig. \ref{fig:PP_OnThetaEpsilon}A), or sometimes only increases the magnitude of density fluctuations (e.g., Fig. \ref{fig:PP_OnThetaEpsilon}B).

\begin{figure}[H]
\centering \includegraphics[scale=0.75, trim = {2cm 1cm 2cm 0}, clip]{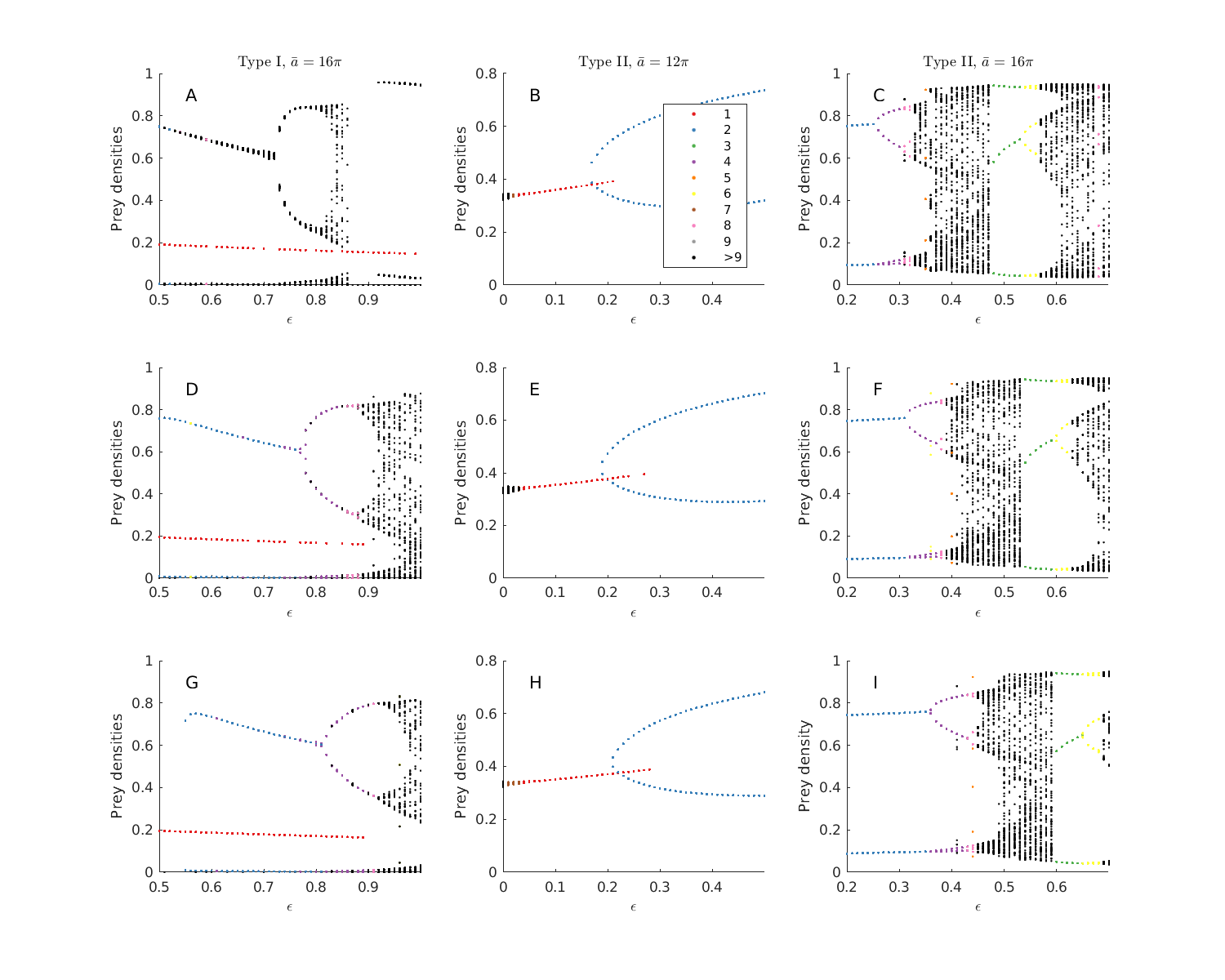}
\caption{Close--ups on bifurcation diagrams on $\epsilon$ for different shapes of the forcing signal and functional responses (left column: type I functional response with $\bar{a} = 16\pi$; middle column: type II functional response with $\bar{a} = 12\pi$; right column: type II functional response with $\bar{a} = 16\pi$). A--B--C) $\theta = 0$ (rectangular signal), D--E--F) $\theta = 0.5$ (intermediate signal), G--H--I) $\theta = 1$ (sinusoidal signal). Colours code for the periodicity of the dynamics which we identify within a range of 1--9 years beyond which we assume quasi--periodic or chaotic regime.}
\label{fig:BifurcDiagEpsilonTheta_Prey}
\end{figure}

However, for higher mean discovery rates, tuning $\theta$ so the transition between summer and winter is more abrupt can considerably change the dynamical behaviour of the community at specific points in the parameter space (Fig. \ref{fig:PP_OnThetaEpsilon}C to F). Figure \ref{fig:BifurcDiagEpsilonTheta_Prey} displays portions of the bifurcation diagrams from Fig. \ref{fig:BifurcDiagOnepsilon_Prey} with period doubling bifurcations for both types of functional responses, and fold bifurcations for the type II functional response (middle and right columns of Fig. \ref{fig:BifurcDiagOnepsilon_Prey}).
Both types of bifurcations occur at lower values of $\epsilon$ when $\theta$ is low for both functional responses (top graphs). For period--doubling bifurcations, this means that increasing the abruptness of seasonal transition produces cycles with periods twice larger for similar seasonal forcing magnitudes. Consequently, the strange attractor (displayed on stroboscopic maps in Fig. \ref{fig:SM_ThetaEpsilon_Type1} and \ref{fig:SM_ThetaEpsilon_Type2}) resulting from multiple period--doublings appears at lower amplitudes $\epsilon$ of the seasonal forcing. For the fold bifurcations, this means that the region of attractor coexistence in Fig. \ref{fig:BifurcDiagEpsilonTheta_Prey}B--E--H is moved to lower values of $\epsilon$, and that dynamics revert to order at lower forcing strengths in Fig. \ref{fig:BifurcDiagEpsilonTheta_Prey}C--F--I. When the forcing signal is rectangular and the functional response is type I (Fig. \ref{fig:BifurcDiagEpsilonTheta_Prey}A), we also detect dynamical behaviours which are not expected with a sinusoidal signal for these parameter values: when $\bar{a} = 16\pi$, most trajectories (aside from the annual cycle in red) are quasi--periodic although they are sometimes near purely periodic orbits (about two years); in addition, the strange attractor (in blue in Fig. \ref{fig:SM_ThetaEpsilon_Type1}A and B) gives way to the annual cycle (Fig. \ref{fig:SM_ThetaEpsilon_Type1}C, D and E) for a short range of $\epsilon$, and then a quasiperiodic behaviour (Fig. \ref{fig:SM_ThetaEpsilon_Type1} F to H) appears and coexists with the annual cycle (Fig. \ref{fig:BifurcDiagEpsilonTheta_Prey}A).

\section{Controlling the variance of the seasonal forcing}
\label{sec:CorrVar}
\subsection{Implementation of the control}

\begin{figure}[H]
\centering \includegraphics[scale = 0.9, trim = {1cm 0.5cm 1cm 0.75cm}, clip]{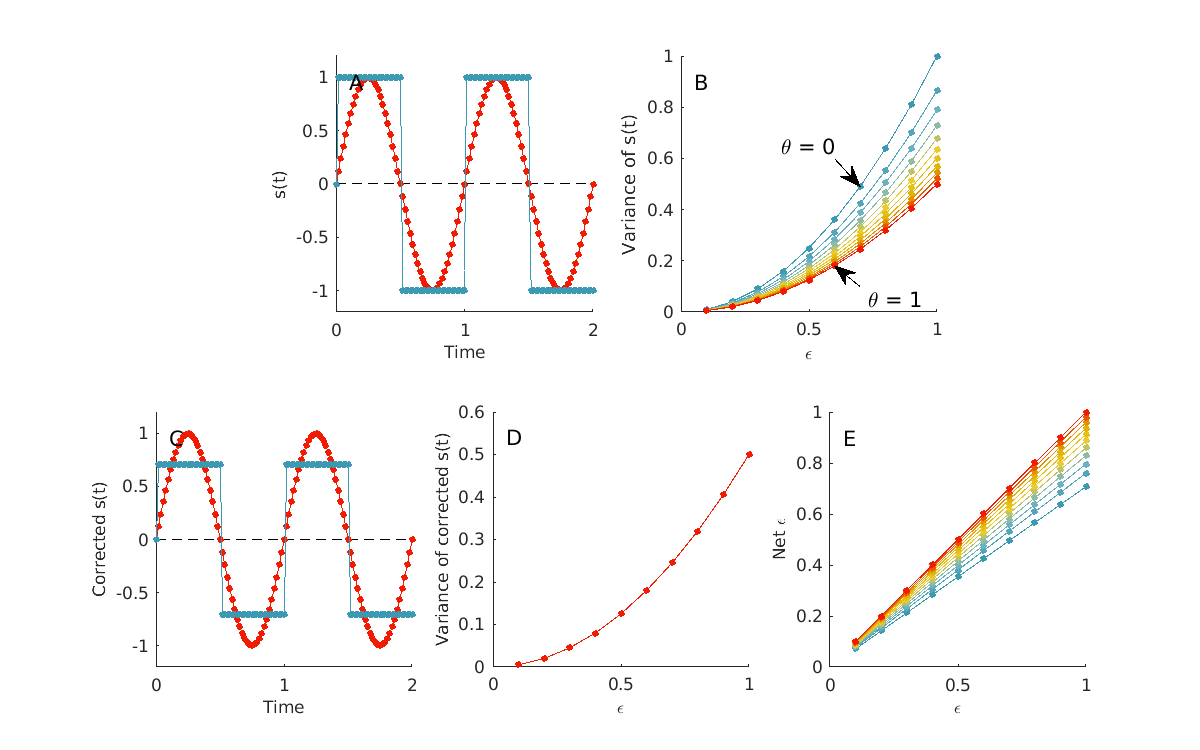}
\caption{Effects of the shape of the forcing signal on its variance over one period. A) Forcing signal $s(t, \epsilon, \theta)$ as described in eq. \eqref{eq:period_func}; B) variance of $s(t, \epsilon, \theta)$; C) forcing signal $\tilde{s}(t, \epsilon, \theta)$ as described in eq. \eqref{eq:period_func_corr}; D) variance of $\tilde{s}(t, \epsilon, \theta)$. As all lines and points overlap, only the variance of $\tilde{s}(t, \epsilon, 1)$ is visible. E) Net magnitude of the seasonal forcing $\tilde{\epsilon}(\theta, \epsilon)$ as $\epsilon \times \nicefrac{\sigma_{q}(1)}{\sigma_{q}(\theta)}$.}
\label{fig:SignalForcingCorr} 
\end{figure}

The shape of the seasonal forcing alters the instantaneous per--capita intake of prey by the predator: when the signal is rectangular ($\theta = 0$), $a(t, \epsilon, \theta)$ is larger for a longer period of time during the predation season (when $s(t, \epsilon, \theta) > 0$) and lower for a longer period of time when $s(t, \epsilon, \theta) < 0$, compared to when the signal is sinusoidal ($\theta = 1$, Fig. \ref{fig:SignalForcingCorr}A).
This translates into differences in the variance of the signal (Fig. \ref{fig:SignalForcingCorr}B), and could explain why cycles can have a larger magnitude when the signal is abrupt and why some bifurcations are crossed at lower magnitude of seasonal forcing.

To test whether differences in the variance of the forcing signal generate the above results, we control the variance of the forcing signal applied on the discovery rate. We do so by correcting $s(t, \epsilon, \theta)$ so its variance no longer depends on $\theta$:

\begin{equation}
\begin{array}{l}
\tilde{s}(t, \epsilon, \theta) = \epsilon \frac{\sigma_{q}(1)}{\sigma_{q}(\theta)}q(t, \theta) \\
q(t, \theta) = \frac{sin(2 \pi t)}{|sin(2 \pi t)|} \times |sin(2 \pi t)|^\theta
\end{array}
\label{eq:period_func_corr}
\end{equation}

\noindent where $\tilde{s}(t, \epsilon, \theta)$ is the corrected forcing signal, $\sigma_{q}(\theta)$ is the standard deviation of $q(t, \theta)$ over one period and $\sigma_{q}(1)$ is the standard deviation of $q(t, \theta)$ when the signal is sinusoidal (i.e., $\theta = 1$).

This approach preserves the bounds of the forcing signal so that $\tilde{s}(t)$ belongs to $[-1, 1]$ despite corrections (Fig. \ref{fig:SignalForcingCorr}C), and ensures identical variances for all $\theta$ while preserving the shape of the seasonal forcing (Fig. \ref{fig:SignalForcingCorr}C--D).

\subsection{Dynamical response to the shape of seasonal forcing when variance is controlled}

Herein, we repeat the figures of section \ref{sec:EpsilonTheta} with the forcing signal corrected following eq. \eqref{eq:period_func_corr}.

\begin{figure}[H]
\centering \includegraphics[scale=0.75, trim = {1.5cm 1cm 2cm 0}, clip]{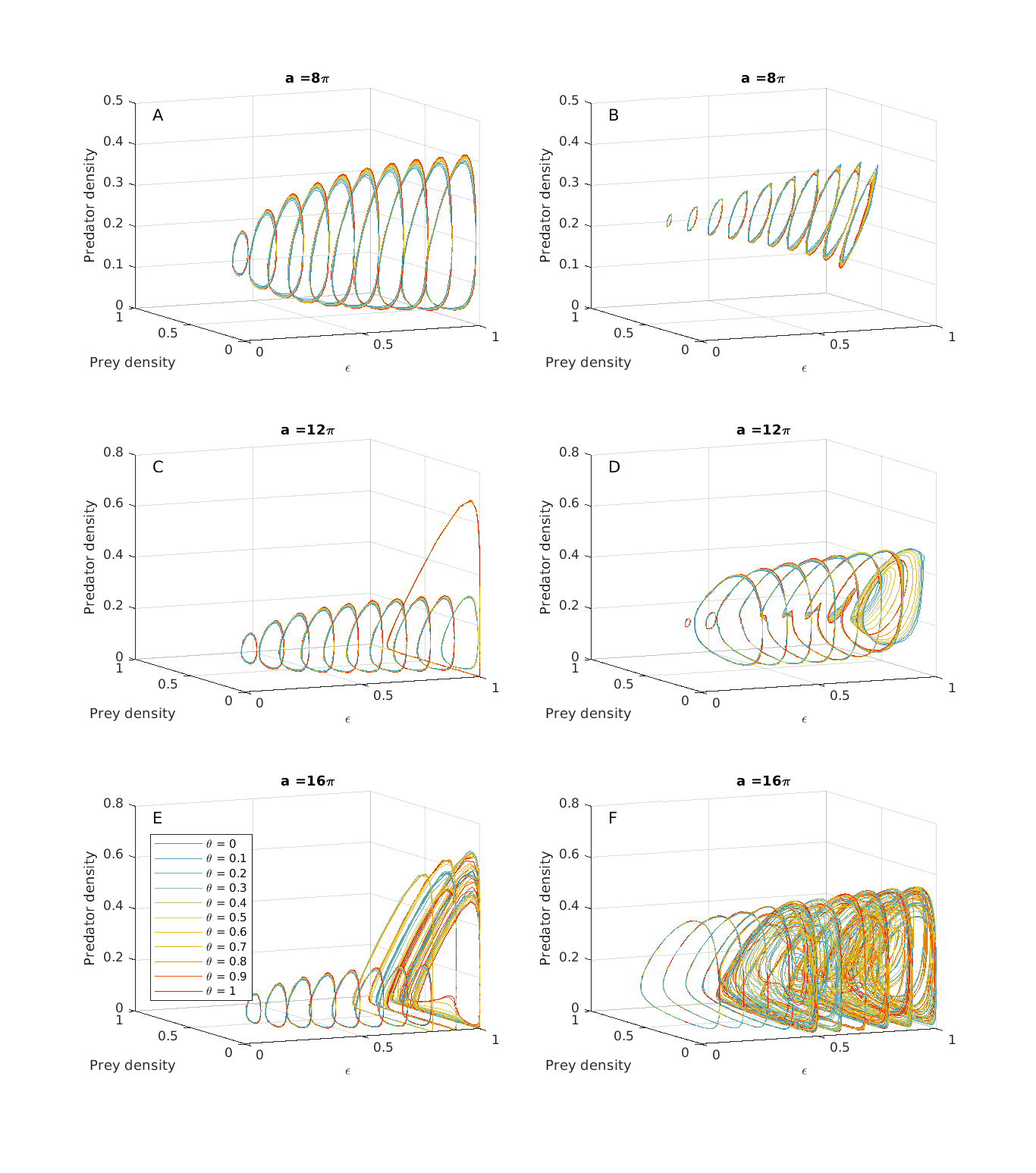}
\caption{Phase portraits of the predator--prey community dynamics undergoing seasonal forcing of various magnitudes $\epsilon$, and abruptness $\theta$, with variance controlled.
Predation is either shaped by a type I functional response (left column: A, C and E), or by a type II functional response (right column: B, D and F).
We consider different mean discovery rates: A--B) $\bar{a}=8\pi$, C--D) $\bar{a}=12\pi$, and E--F) $\bar{a}=16\pi$.
Simulations are all initiated with the same densities $(x_0,y_0) = (0.3,0.3)$.
 Only asymptotic dynamics covering the last ten years of simulation are displayed.
 Closed curves correspond to periodic dynamics while other attractors are indicative of quasi--periodic and chaotic behaviours.}
\label{fig:PP_OnThetaEpsilon_Corr} 
\end{figure}

When we fix the mean discovery rate $\bar{a}$ such that no bifurcations are crossed when $\epsilon$ is varied ($\bar{a} = 8\pi$), controlling the variance of the forcing signal greatly reduces differences between community trajectories for different values of $\theta$, although density fluctuations may still have different magnitudes (Fig. \ref{fig:PP_OnThetaEpsilon_Corr}A and B). Hence, the variance of the signal is not sufficient to explain the changes in the magnitude of population fluctuations when the signal gets more rectangular.

\begin{figure}[H]
\centering
\includegraphics[scale=0.75, trim = {2cm 1cm 2cm 0}, clip]{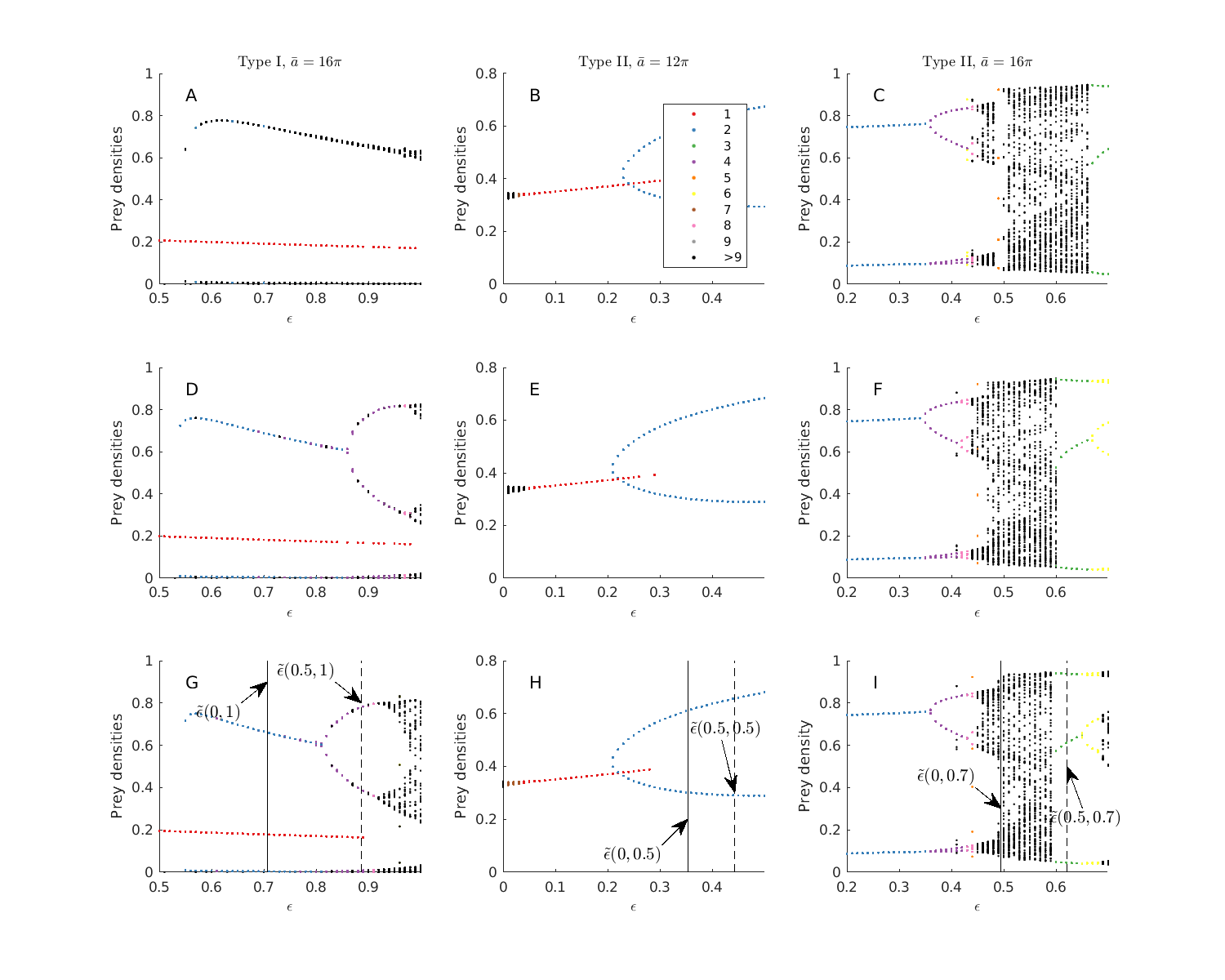}
\caption{Close--ups on the bifurcation diagram on $\epsilon$ for different shapes of the forcing signal when its variance is controlled (left column: type I functional response with $\bar{a} = 16\pi$; middle column: type II functional response with $\bar{a} = 12\pi$; right column: type II functional response with $\bar{a} = 16\pi$): A--B--C) $\theta = 0$ (rectangular signal), D--E--F) $\theta = 0.5$ (intermediate signal), G--H--I) $\theta = 1$ (sinusoidal signal). Colours code for the periodicity of the dynamics which we identify within a range of 1--9 years beyond which we assume quasi--periodic or chaotic regime. Vertical lines drawn in G, H and I indicate the net forcing magnitude $\tilde{\epsilon}(\theta, \epsilon) = \epsilon \frac{\sigma_q(1)}{\sigma_q(\theta)}$ corresponding to the upper bounds of the bifurcation diagrams A to F for which $\theta < 1$ (solid lines for $\theta = 0$, first row maximum $\tilde{\epsilon}$; dashed line for $\theta = 0.5$, second row maximum $\tilde{\epsilon}$).}
\label{fig:BifurcDiagEpsilonTheta_CV_Prey} 
\end{figure}

Turning to higher values of the mean discovery rate $\bar{a}$, we explore how controlling the variance of the signal affects the sequence of bifurcations with respect to $\epsilon$. There is no early bifurcation anymore when the signal gets more rectangular (Fig. \ref{fig:BifurcDiagEpsilonTheta_CV_Prey}). In fact, the relationship between $\theta$ and the value of $\epsilon$ at which bifurcations occur is reversed (Fig. \ref{fig:PP_OnThetaEpsilon_Corr}B to F, Fig. \ref{fig:BifurcDiagEpsilonTheta_CV_Prey}). Namely, when the functional response of the predator is of type I, period doubling occurs at a higher strength of seasonal forcing when the forcing signal tends towards rectangular than when sinusoidal (Fig. \ref{fig:BifurcDiagEpsilonTheta_CV_Prey}D and G). When the signal is fully rectangular, the two--year cycle even seems to turn directly into a quasi--periodic attractor (Fig. \ref{fig:BifurcDiagEpsilonTheta_CV_Prey}A). An indirect consequence of period--doubling bifurcations occurring at higher values of seasonal forcing is that the system reaches the strange attractor, characterising community dynamics at high seasonal forcing, for a higher value of $\epsilon$ (illustrated in Fig. \ref{fig:SM_CV_ThetaEpsilon_Type1} for the type I functional response). Similarly, when the functional response of the predator is of type II, both the fold bifurcation and the period--doubling bifurcation delineating the region of attractors' coexistence move to higher values of $\epsilon$ when the signal gets more rectangular (Fig. \ref{fig:BifurcDiagEpsilonTheta_CV_Prey}B, E, H, and Fig. \ref{fig:SM_CV_ThetaEpsilon_Type2}). For $\bar{a} = 16\pi$, as period--doubling bifurcations and fold bifurcations occur at higher $\epsilon$ for more rectangular forcing signals, the dynamics return to order for higher forcing strength, and the second period--doubling route to chaos starting on three--year cycles is no longer observed within the  considered range of $\epsilon$--values (Fig. \ref{fig:BifurcDiagEpsilonTheta_Prey}C, F and I).

As we control the variance of the forcing signal, we simultaneously reduce the net magnitude of the forcing exerted on the discovery rate ($\tilde{\epsilon}(\theta, \epsilon) = \epsilon \frac{\sigma_{q}(1)}{\sigma_{q}(\theta)}$ in eq. \eqref{eq:period_func_corr}, Fig. \ref{fig:SignalForcingCorr}E). For instance, this is equivalent to sampling the bifurcation diagram of Fig. \ref{fig:BifurcDiagEpsilonTheta_CV_Prey}G between $\tilde{\epsilon}(\theta, 0.5) = 0.5 \times \frac{\sigma_{q}(1)}{\sigma_{q}(\theta)}$ and $\tilde{\epsilon}(\theta, 1) = \frac{\sigma_{q}(1)}{\sigma_{q}(\theta)}$ (see Fig. \ref{fig:BifurcDiagEpsilonTheta_CV_Prey}G, H and I where upper bounds of these bifurcation subsets for $\theta = \{0,\,0.5\}$ are drawn with vertical lines). Late bifurcations could thus be thought to be a side--product of decreasing the magnitude of parameter fluctuations. For instance, prey density at the beginning of each year in Fig. \ref{fig:BifurcDiagEpsilonTheta_CV_Prey}A is very similar to the corresponding subset in Fig. \ref{fig:BifurcDiagEpsilonTheta_CV_Prey}G (to the left of the vertical solid line). However attractors periodicities do not fully match as one attractor is quasi--periodic when $\theta = 0$ while its counterpart is a two--year cycle when $\theta = 1$. When $\bar{a} = 16\pi$ and the functional response of type II, differences between attractors produced by rectangular signals and attractors produced by the sinusoidal signal are more striking. For example, if Fig. \ref{fig:BifurcDiagEpsilonTheta_CV_Prey}C were equivalent to the left subset of Fig. \ref{fig:BifurcDiagEpsilonTheta_CV_Prey}I (on the left of the solid line), we would not observe the fold bifurcation allowing chaotic regime to turn into a three--year cycle.

These results therefore indicate that neither the variance nor the strength of the forcing signal are sufficient to predict the dynamical behaviour of the community. Other features of the seasonal forcing, which determine the amount of forcing occurring within one predation season, are also affecting the timing of bifurcations.

\section{Discussion}

Seasonal forcing on ecological processes can take more or less abrupt shapes depending on the driving environmental variable (Fig. \ref{fig:ClimateForcing}).
Yet, most theoretical studies on the dynamical consequences of seasonal forcing essentially focus on a sinusoidal forcing, assuming all ecological parameters follow the wave--like pattern of temperature. Besides, little to no connection is made with studies investigating more realistic forcing (e.g., abrupt changes of habitat features affecting species mimicry).
Consequently, whether the type of seasonal transition (smooth or abrupt) of 
a parameter value affects predator--prey dynamics remains unclear, and the generality of theory to different seasonal transitions needs to be assessed.
To address this, we have explored the dynamical consequences of seasonal 
predation through numerical simulations of predator--prey communities when 
the predator's discovery rate is a periodic function of time. We considered 
jointly two properties of the seasonal forcing signal, namely the magnitude 
of parameter fluctuations $\epsilon$ and the abruptness of the forcing 
signal $\theta$, to assess how the shape of seasonal transition affects 
community dynamics.
We found that abrupt seasonal transitions trigger bifurcations at lower forcing strengths, hence producing dynamical behaviours which are not expected in some parameter regions. In addition, we showed that the variance of the forcing signal is not sufficient to explain where bifurcations lie on the $\epsilon$--axis, suggesting that other features of the forcing signal drive the dynamical behaviour of the system. To provide some context to our results, we first review the dynamical behaviour generated by seasonal forcing on predator--prey models, and then discuss the dynamical consequences of more abrupt forcing signals.

Forcing the predator's discovery rate aims at mimicking seasonal fluctuations in predators' ability to detect and kill prey on a given site. Indeed, predation success can vary across seasons because species' ability -- whether prey or predator, to hide or move (to hunt or escape) in their habitat fluctuates with environmental conditions. This is equivalent to the fifth mechanism listed by \citet{rinaldi_multiple_1993}, which only alters the rate of increase of predator functional response (type I functional response), and how fast it saturates \citep[type II functional response,][]{kuznetsov_bifurcations_1992}. In the latter case, our model actually becomes the Rosenzweig--MacArthur model (see Supplementary information) which has been examined under seasonal forcing in many theoretical studies \citep[e.g.,][]{kuznetsov_bifurcations_1992,rinaldi_multiple_1993,king_rainbow_1999,king_geometry_2001,taylor_seasonal_2013,taylor_how_2013}. Works on seasonal predator--prey communities governed by a linear functional response are rarer \citep[e.g.,][]{inoue_scenarios_1984,vandermeer_categories_2001,bibik_investigation_2015} and do not necessarily involve logistic growth of prey as in our model \citep[but see][]{sabin_chaos_1993,king_weakly_1996,vandermeer_seasonal_1996}. For most studies on seasonal forcing, the focus is on prey reproduction as a primary entry for seasonality in food webs \citep[but see][]{stollenwerk_hopf_2017}, assuming predators solely adjust their reproduction to prey density and not necessarily their effective availability. However, \citet{rinaldi_multiple_1993} found that a universal bifurcation diagram can be drawn for the Rosenzweig--MacArthur model for all  parameters when subjected to seasonal forcing, provided that they are bifurcation parameters in the unforced system \citep{taylor_seasonal_2013}. Hence, despite differences with our models, these earlier theoretical studies enable us to clarify the outcome of our simulations.

Before altering the shape of seasonal forcing beyond standard sinusoidal, we have examined the dynamical behaviour of the seasonally forced model considering two types of functional responses of the predator. The most common dynamical behaviour produced by periodic forcing of the discovery rate is a cycle with the same period as the forcing signal. When considering a saturating functional response, this behaviour occurs when the discovery rate of the unforced system lies below the Hopf bifurcation \citep[where the stable point becomes a limit cycle;][]{rinaldi_multiple_1993,taylor_seasonal_2013}. Increasing the magnitude of the forcing on the predator's discovery rate brings about period doubling bifurcations for both types of functional responses. In the case of linear intake of prey, the multiple period--doubling bifurcations follow one after the other before the dynamics become chaotic \citep{vandermeer_seasonal_1996}, i.e., there is a period--doubling route to chaos. The Rosenzweig--MacArthur model can fall into a chaotic regime in more diverse ways for the same value of $\bar{a}$ (the mean discovery rate) provided that the unforced dynamics are already cyclic \citep{rinaldi_multiple_1993,rinaldi_conditioned_1993}. However, our parameter choice does not allow observation of the torus destruction route to chaos, as our system always enters chaotic regions through multiple period--doubling bifurcations, and revert to order through fold bifurcations \citep{rinaldi_multiple_1993}.

In addition, we were able to detect a range for $\epsilon$ where coexistence of annual cycles with multi--annual cycles is possible. These coexisting attractors are very widespread in conservative (Hamiltonian) predator--prey models \citep{inoue_scenarios_1984,vandermeer_categories_2001,bibik_investigation_2015}. However, many ecological systems of interest are dissipative \citep{turchin_population_1995}. Regarding this more realistic case, \citet{king_weakly_1996} showed that coexistence of multiple attractors with various periods for systems driven by a linear functional response of predators is possible even if the prey's growth rate is bounded. The number of alternative attractors nonetheless decreases when the dissipation level of the system ($\nicefrac{d}{Kac}$) increases which could explain the relatively small number of coexisting cycles we observe. Indeed, although the predator's discovery rate is high enough to allow species coexistence ($\nicefrac{d}{Kac} < 1$), our parameterisation corresponds to a rather dissipative system. This also goes in hand with the rather high forcing strength necessary to observe the coexisting cycles in our simulations. Coexistence of periodic solutions exists in the Rosenzweig--MacArthur model as well, and they are contained within a parameter area delineated by a Neimark--Sacker bifurcation, a fold bifurcation, and a period--doubling bifurcation \citep{rinaldi_multiple_1993,kuznetsov_bifurcations_1992}. The presence of overlapping Arnol'd tongues is another explanation of the coexistence of these different dynamical regimes \citep{greenman_large_2004}. Arnol'd tongues are parameter regions where the system displays cyclic behaviours whose periodicities are rational numbers of the period of the forcing signal. They are usually rooted in Neimark--Sacker bifurcations and framed by fold bifurcations \citep{dercole_dynamical_2011,taylor_seasonal_2013}. We have thus encountered some of these parameter regions in our numerical simulations (see periodic windows in Fig. \ref{fig:BifurcDiagOnepsilon_Prey}F).

By tuning the shape of the seasonal forcing so that it ranges between fully sinusoidal to fully rectangular, we found that more abrupt seasonal transitions in the value of discovery rate can result in very different outcomes: at low discovery rates, although the community trajectory may remain roughly unchanged, density fluctuations can become larger, while early bifurcations can happen at higher discovery rates. In this latter case, this means that for the same magnitude of seasonal forcing (our $\epsilon$) -- implying the same year--round average discovery rate, the seasonally forced community can produce more complex dynamics with abrupt forcing than expected with sinusoidal forcing. It also echoes \citet{wiesenfeld_noisy_1985}'s detection of bifurcation precursors through stochastic forcing. The early bifurcations we found align with \citet{turchin_empirically_1997}'s observation of more chaotic dynamics when considering a rectangular seasonality function. However, they did not observe increased population fluctuations, which may be due to their focus on their empirically--based parameterisation. Overall, our results suggest that the bifurcation landscape may be modified by the parameter $\theta$, making it a bifurcation parameter.

To our knowledge, few studies actually question the robustness of their results to the shape of the seasonal transitions, but of those that do most are epidemiological models \citep{greenman_external_2004,ireland_effect_2004}. The shape of seasonality functions they explored, in addition to the common sinusoidal forcing, is diverse as it includes both rectangular and saw--tooth signals. Interestingly, they report (without showing simulations) that the above--mentioned effects of more abrupt seasonal forcing are lost when controlling for the variance of the signal. Indeed, different variances of the forcing signal can be interpreted in physical terms as the energy of the forcing. Building on their observation, we further investigated this lead and found that controlling the variance of the forcing signal is actually not sufficient to dim all differences between trajectories. Not only does our correction of the signal variance prevent early bifurcations and early emergence of chaos but it even reverses the consequences of abrupt seasonal forcing by postponing bifurcations until a higher forcing magnitude, when the signal is more rectangular. In order for our correction to control the variance of the signal and preserve its shape, the net strength of seasonal forcing is reduced for more rectangular signals, and this could partly explain why some dynamical behaviours are postponed to larger forcing amplitudes. With this correction, one should actually contrast the bifurcation diagrams with $\theta < 1$ with subsets of the bifurcation diagram with $\theta = 1$ (fully sinusoidal forcing) for a more relevant comparison. As we argued in section \ref{sec:CorrVar}, this is because the range of net $\epsilon$ ($\epsilon \times \nicefrac{\sigma_{q}(1)}{\sigma_{q}(\theta)}$ in eq. \eqref{eq:period_func_corr}) covered is actually smaller than $[0,\,1]$ (see Fig. \ref{fig:SignalForcingCorr}E). Despite this, some differences of dynamical behaviours persist, suggesting that the dynamical response to seasonal forcing truly depends on the shape of the forcing signal. In other words, a forcing signal cannot be considered as solely a variance mechanism or an amplitude effect: multiple aspects of a forcing signal act together on the dynamics of a system.

Here, we hypothesise that the shape of the forcing signal could actually interact with species densities over time. Indeed, during the predation season ($s(t, \epsilon, \theta) > 0$), the per--capita intake of prey by the predator is higher and more constant for a longer period of time when a signal is rectangular than when the signal is sinusoidal, even if variances are comparable. This allows a more consistent increase (decrease) of predator (prey) density during this season. Similarly, the per--capita intake will be lower for a longer period of time during the resting season ($s(t, \epsilon, \theta) < 0$) when the signal is rectangular, which allows the prey to recover more consistently. This type of mechanism could explain why, despite controlled variance, the magnitude of density fluctuations remains larger when the forcing signal tends towards rectangularity. Hence, the effect of the shape of the forcing signal might also be the result of an interaction between the instantaneous per capita intake of the predator (the sequence of $s(t, \theta, \epsilon)$) and species densities.

\section{Conclusion}

Can seasonal forcing be reduced to the amplitude and periodicity of a sine wave? Our simulations of the dynamics of predator--prey communities tell a more balanced story. As implementing more abrupt seasonal shifts leads community dynamics into bifurcations and chaotic dynamics for lower seasonal forcing strengths, the shape of seasonal forcing stands out as a contributing driver to the fluctuations in densities and the type of dynamical behaviour observed. Contrary to previous works, we found that this effect is not a mere consequence of the variance of the forcing signal, even though increasing the variance of the forcing signal does help, all other things being equal, to cross bifurcations more easily. This should motivate future work to include in models of community dynamics more lifelike seasonal temporal variation in parameters.
More broadly, characterising how environmental fluctuations translate into temporal variation in species life history schedules and foraging behaviours may help to better predict the temporal dynamics of ecological communities.

\section*{Acknowledgements}
This work was funded by the French ANR through LabEx COTE (ANR--10--LABX--45). We thank Sébastien Lafont for support on handling CDS data  and an anonymous reviewer for valuable input.

\section*{Data availability}
Computer codes for simulations are available in GitHub repository \href{https://github.com/alixsauve/ShapeOfSeasonality}{alixsauve/ShapeOfSeasonality} and have been archived at Zenodo with \href{https://doi.org/10.5281/zenodo.3687867}{https://doi.org/10.5281/zenodo.3687867}. 

\pagebreak{}

\bibliographystyle{elsarticle-harv}
\bibliography{RefList2SpSeason}

\begin{thebibliography}{48}
\expandafter\ifx\csname natexlab\endcsname\relax\def\natexlab#1{#1}\fi
\expandafter\ifx\csname url\endcsname\relax
  \def\url#1{\texttt{#1}}\fi
\expandafter\ifx\csname urlprefix\endcsname\relax\def\urlprefix{URL }\fi

\bibitem[{Barraquand et~al.(2017)Barraquand, Louca, Abbott, Cobbold,
  Cordoleani, DeAngelis, Elderd, Fox, Greenwood, Hilker, and
  {others}}]{barraquand_moving_2017}
Barraquand, F., Louca, S., Abbott, K.~C., Cobbold, C.~A., Cordoleani, F.,
  DeAngelis, D.~L., Elderd, B.~D., Fox, J.~W., Greenwood, P., Hilker, F.~M.,
  {others}, 2017. Moving forward in circles: challenges and opportunities in
  modelling population cycles. Ecology letters 20~(8), 1074--1092.

\bibitem[{Bibik(2015)}]{bibik_investigation_2015}
Bibik, Y.~V., 2015. Investigation of transition to chaos for a
  {Lotka}–{Volterra} system with the seasonality factor using the dissipative
  {Henon} map. Applied Mathematical Sciences 9~(117), 5801--5837.

\bibitem[{Bilodeau et~al.(2013)Bilodeau, Gauthier, and
  Berteaux}]{bilodeau_effect_2013}
Bilodeau, F., Gauthier, G., Berteaux, D., 2013. Effect of snow cover on the
  vulnerability of lemmings to mammalian predators in the {Canadian} {Arctic}.
  Journal of Mammalogy 94~(4), 813--819.

\bibitem[{Bjørnstad et~al.(1995)Bjørnstad, Falck, and
  Stenseth}]{bjornstad_geographic_1995}
Bjørnstad, O.~N., Falck, W., Stenseth, N.~C., 1995. A geographic gradient in
  small rodent density fluctuations: a statistical modelling approach.
  Proceedings of the Royal Society of London. Series B: Biological Sciences
  262~(1364), 127--133.

\bibitem[{{Copernicus Climate Change Service (C3S)}(2017)}]{era5}
{Copernicus Climate Change Service (C3S)}, 2017. {ERA}5: Fifth generation of
  {ECMWF} atmospheric reanalyses of the global climate. Copernicus Climate
  Change Service Climate Data Store (CDS).
  \url{https://cds.climate.copernicus.eu/cdsapp}, [Online; accessed
  13-December-2019].

\bibitem[{Dercole and Rinaldi(2011)}]{dercole_dynamical_2011}
Dercole, F., Rinaldi, S., 2011. Dynamical systems and their bifurcations.
  Advanced Methods of Biomedical Signal Processing, 291--325.

\bibitem[{Dhooge et~al.(2003)Dhooge, Govaerts, and
  Kuznetsov}]{dhooge_matcont:_2003}
Dhooge, A., Govaerts, W., Kuznetsov, Y.~A., 2003. {MATCONT}: a {MATLAB} package
  for numerical bifurcation analysis of {ODEs}. ACM Transactions on
  Mathematical Software (TOMS) 29~(2), 141--164.

\bibitem[{Doedel(1981)}]{doedel_auto:_1981}
Doedel, E.~J., 1981. {AUTO}: {A} program for the automatic bifurcation analysis
  of autonomous systems. Congressus Numerantium 30~(265-284), 25--93.

\bibitem[{{ECMWF}(2016)}]{ecmwf_part_2016}
{ECMWF}, 2016. Part {IV}: {P}hysical {P}rocesses. In: {IFS} {D}ocumentation
  {CY}43R1. No.~4 in {IFS} {D}ocumentation. ECMWF.
\newline\urlprefix\url{https://www.ecmwf.int/node/17117}

\bibitem[{Fauteux et~al.(2015)Fauteux, Gauthier, and
  Berteaux}]{fauteux_seasonal_2015}
Fauteux, D., Gauthier, G., Berteaux, D., 2015. Seasonal demography of a cyclic
  lemming population in the {Canadian} {Arctic}. Journal of Animal Ecology
  84~(5), 1412--1422.

\bibitem[{Gilg et~al.(2003)Gilg, Hanski, and Sittler}]{gilg_cyclic_2003}
Gilg, O., Hanski, I., Sittler, B., 2003. Cyclic dynamics in a simple vertebrate
  predator-prey community. Science 302~(5646), 866--868.

\bibitem[{Gragnani and Rinaldi(1995)}]{gragnani_universal_1995}
Gragnani, A., Rinaldi, S., 1995. A universal bifurcation diagram for seasonally
  perturbed predator-prey models. Bulletin of Mathematical Biology 57~(5),
  701--712.

\bibitem[{Greenman and Benton(2004)}]{greenman_large_2004}
Greenman, J., Benton, T., 2004. Large amplification in stage-structured models:
  {Arnol}'d tongues revisited. Journal of Mathematical Biology 48~(6),
  647--671.

\bibitem[{Greenman et~al.(2004)Greenman, Kamo, and
  Boots}]{greenman_external_2004}
Greenman, J., Kamo, M., Boots, M., 2004. External forcing of ecological and
  epidemiological systems: a resonance approach. Physica D: Nonlinear Phenomena
  190~(1-2), 136--151.

\bibitem[{Hanski et~al.(1993)Hanski, Turchin, Korpimäki, and
  Henttonen}]{hanski_population_1993}
Hanski, I., Turchin, P., Korpimäki, E., Henttonen, H., 1993. Population
  oscillations of boreal rodents: regulation by mustelid predators leads to
  chaos. Nature 364~(6434), 232.

\bibitem[{Hansson and Henttonen(1985)}]{hansson_gradients_1985}
Hansson, L., Henttonen, H., 1985. Gradients in density variations of small
  rodents: the importance of latitude and snow cover. Oecologia 67, 394--402.

\bibitem[{Inoue and Kamifukumoto(1984)}]{inoue_scenarios_1984}
Inoue, M., Kamifukumoto, H., 1984. Scenarios leading to chaos in a forced
  {Lotka}-{Volterra} model. Progress of Theoretical Physics 71~(5), 930--937.

\bibitem[{Ireland et~al.(2004)Ireland, Norman, and
  Greenman}]{ireland_effect_2004}
Ireland, J.~M., Norman, R., Greenman, J., 2004. The effect of seasonal host
  birth rates on population dynamics: the importance of resonance. Journal of
  Theoretical Biology 231~(2), 229--238.

\bibitem[{King et~al.(1996)King, Schaffer, Gordon, Treat, and
  Kot}]{king_weakly_1996}
King, A., Schaffer, W.~M., Gordon, C., Treat, J., Kot, M., 1996. Weakly
  dissipative predator-prey systems. Bulletin of Mathematical Biology 58~(5),
  835--859.

\bibitem[{King and Schaffer(1999)}]{king_rainbow_1999}
King, A.~A., Schaffer, W.~M., 1999. The rainbow bridge: {Hamiltonian} limits
  and resonance in predator-prey dynamics. Journal of Mathematical Biology
  39~(5), 439--469.

\bibitem[{King and Schaffer(2001)}]{king_geometry_2001}
King, A.~A., Schaffer, W.~M., 2001. The geometry of a population cycle: {A}
  mechanistic model of snowshoe hare demography. Ecology 82~(3), 814--830.

\bibitem[{Korslund and Steen(2006)}]{korslund_small_2006}
Korslund, L., Steen, H., 2006. Small rodent winter survival: snow conditions
  limit access to food resources. Journal of Animal Ecology 75~(1), 156--166.

\bibitem[{Kuang(1988)}]{kuang_nonuniqueness_1988}
Kuang, Y., 1988. Nonuniqueness of limit cycles of {Gause}-type predator-prey
  systems. Applicable Analysis 29~(3-4), 269--287.

\bibitem[{Kuznetsov et~al.(1992)Kuznetsov, Muratori, and
  Rinaldi}]{kuznetsov_bifurcations_1992}
Kuznetsov, Y.~A., Muratori, S., Rinaldi, S., 1992. Bifurcations and chaos in a
  periodic predator-prey model. International Journal of Bifurcation and Chaos
  2~(01), 117--128.

\bibitem[{Magnhagen(1991)}]{magnhagen_predation_1991}
Magnhagen, C., 1991. Predation risk as a cost of reproduction. Trends in
  Ecology \& Evolution 6~(6), 183--186.

\bibitem[{Mott et~al.(2018)Mott, Vionnet, and Grünewald}]{mott_seasonal_2018}
Mott, R., Vionnet, V., Grünewald, T., 2018. The seasonal snow cover dynamics:
  review on wind-driven coupling processes. Frontiers in Earth Science 6, 197.

\bibitem[{Pau et~al.(2011)Pau, Wolkovich, Cook, Davies, Kraft, Bolmgren,
  Betancourt, and Cleland}]{pau_predicting_2011}
Pau, S., Wolkovich, E.~M., Cook, B.~I., Davies, T.~J., Kraft, N.~J., Bolmgren,
  K., Betancourt, J.~L., Cleland, E.~E., 2011. Predicting phenology by
  integrating ecology, evolution and climate science. Global Change Biology
  17~(12), 3633--3643.

\bibitem[{Penczykowski et~al.(2017)Penczykowski, Connolly, and
  Barton}]{penczykowski_winter_2017}
Penczykowski, R.~M., Connolly, B.~M., Barton, B.~T., Dec. 2017. Winter is
  changing: {Trophic} interactions under altered snow regimes. Food Webs 13,
  80--91.

\bibitem[{Probst et~al.(2011)Probst, Nemeschkal, McGrady, Tucakov, and
  Szép}]{probst_aerial_2011}
Probst, R., Nemeschkal, H., McGrady, M., Tucakov, M., Szép, T., 2011. Aerial
  hunting techniques and predation success of {Hobbies} \emph{Falco subbuteo}
  on {Sand} {Martin} \emph{Riparia riparia} at breeding colonies. Ardea 99~(1),
  9--17.

\bibitem[{Reynolds et~al.(2013)Reynolds, Sherratt, White, and
  Lambin}]{reynolds_comparison_2013}
Reynolds, J. J.~H., Sherratt, J.~A., White, A., Lambin, X., 2013. A comparison
  of the dynamical impact of seasonal mechanisms in a herbivore–plant defence
  system. Theoretical Ecology 6~(2), 225--239.

\bibitem[{Rinaldi and Muratori(1993)}]{rinaldi_conditioned_1993}
Rinaldi, S., Muratori, S., 1993. Conditioned chaos in seasonally perturbed
  predator-prey models. Ecological Modelling 69~(1-2), 79--97.

\bibitem[{Rinaldi et~al.(1993)Rinaldi, Muratori, and
  Kuznetsov}]{rinaldi_multiple_1993}
Rinaldi, S., Muratori, S., Kuznetsov, Y., 1993. Multiple attractors,
  catastrophes and chaos in seasonally perturbed predator-prey communities.
  Bulletin of mathematical Biology 55~(1), 15--35.

\bibitem[{Rosenzweig and MacArthur(1963)}]{rosenzweig_graphical_1963}
Rosenzweig, M.~L., MacArthur, R.~H., 1963. Graphical representation and
  stability conditions of predator-prey interactions. The American Naturalist
  97~(895), 209--223.

\bibitem[{Sabin and Summers(1993)}]{sabin_chaos_1993}
Sabin, G.~C., Summers, D., 1993. Chaos in a periodically forced predator-prey
  ecosystem model. Mathematical Biosciences 113~(1), 91--113.

\bibitem[{Saino et~al.(2010)Saino, Ambrosini, Rubolini, von Hardenberg,
  Provenzale, Hüppop, Hüppop, Lehikoinen, Lehikoinen, Rainio, and
  {others}}]{saino_climate_2010}
Saino, N., Ambrosini, R., Rubolini, D., von Hardenberg, J., Provenzale, A.,
  Hüppop, K., Hüppop, O., Lehikoinen, A., Lehikoinen, E., Rainio, K.,
  {others}, 2010. Climate warming, ecological mismatch at arrival and
  population decline in migratory birds. Proceedings of the Royal Society B:
  Biological Sciences 278~(1707), 835--842.

\bibitem[{Sonerud(1986)}]{sonerud_effect_1986}
Sonerud, G.~A., 1986. Effect of snow cover on seasonal changes in diet,
  habitat, and regional distribution of raptors that prey on small mammals in
  boreal zones of {Fennoscandia}. Ecography 9, 33--47.

\bibitem[{Stenseth et~al.(2004)Stenseth, Shabbar, Chan, Boutin, Rueness,
  Ehrich, Hurrell, Lingjærde, and Jakobsen}]{stenseth_snow_2004}
Stenseth, N.~C., Shabbar, A., Chan, K.-S., Boutin, S., Rueness, E.~K., Ehrich,
  D., Hurrell, J.~W., Lingjærde, O.~C., Jakobsen, K.~S., 2004. Snow conditions
  may create an invisible barrier for lynx. Proceedings of the National Academy
  of Sciences 101~(29), 10632--10634.

\bibitem[{Stollenwerk et~al.(2017)Stollenwerk, Sommer, Kooi, Mateus, Ghaffari,
  and Aguiar}]{stollenwerk_hopf_2017}
Stollenwerk, N., Sommer, P.~F., Kooi, B., Mateus, L., Ghaffari, P., Aguiar, M.,
  2017. Hopf and torus bifurcations, torus destruction and chaos in population
  biology. Ecological Complexity 30, 91--99.

\bibitem[{Taylor et~al.(2013{\natexlab{a}})Taylor, Sherratt, and
  White}]{taylor_seasonal_2013}
Taylor, R.~A., Sherratt, J.~A., White, A., 2013{\natexlab{a}}. Seasonal forcing
  and multi-year cycles in interacting populations: lessons from a
  predator–prey model. Journal of mathematical biology 67~(6-7), 1741--1764.

\bibitem[{Taylor et~al.(2013{\natexlab{b}})Taylor, White, and
  Sherratt}]{taylor_how_2013}
Taylor, R.~A., White, A., Sherratt, J.~A., 2013{\natexlab{b}}. How do
  variations in seasonality affect population cycles? Proceedings of the Royal
  Society B: Biological Sciences 280~(1754), 20122714.

\bibitem[{Therrien et~al.(2014)Therrien, Gauthier, Pinaud, and
  Bêty}]{therrien_irruptive_2014}
Therrien, J.-F., Gauthier, G., Pinaud, D., Bêty, J., 2014. Irruptive movements
  and breeding dispersal of snowy owls: a specialized predator exploiting a
  pulsed resource. Journal of Avian Biology 45~(6), 536--544.

\bibitem[{Tornberg(1997)}]{tornberg_prey_1997}
Tornberg, R., 1997. Prey selection of the goshawk \emph{Accipiter gentilis}
  during the breeding season: the role of prey profitability and vulnerability.
  Ornis Fennica 74~(1), 15--28.

\bibitem[{Turchin(1995)}]{turchin_population_1995}
Turchin, P., 1995. Population regulation: old arguments and a new synthesis.
  In: Cappuccino, N., Price, P.~W. (Eds.), Population dynamics: new approaches
  and synthesis. Academic Press, New York, pp. 19--39.

\bibitem[{Turchin and Hanski(1997)}]{turchin_empirically_1997}
Turchin, P., Hanski, I., 1997. An empirically based model for latitudinal
  gradient in vole population dynamics. The American Naturalist 149~(5),
  842--874.

\bibitem[{Tyson and Lutscher(2016)}]{tyson_seasonally_2016}
Tyson, R., Lutscher, F., 2016. Seasonally varying predation behavior and
  climate shifts are predicted to affect predator-prey cycles. The American
  Naturalist 188~(5), 539--553.

\bibitem[{Vandermeer(1996)}]{vandermeer_seasonal_1996}
Vandermeer, J., 1996. Seasonal isochronic forcing of {Lotka}--{Volterra}
  equations. Progress of Theoretical Physics 96~(1), 13--28.

\bibitem[{Vandermeer et~al.(2001)Vandermeer, Stone, and
  Blasius}]{vandermeer_categories_2001}
Vandermeer, J., Stone, L., Blasius, B., 2001. Categories of chaos and fractal
  basin boundaries in forced predator–prey models. Chaos, Solitons \&
  Fractals 12~(2), 265--276.

\bibitem[{Wiesenfeld(1985)}]{wiesenfeld_noisy_1985}
Wiesenfeld, K., 1985. Noisy precursors of nonlinear instabilities. Journal of
  Statistical Physics 38~(5-6), 1071--1097.

\end{thebibliography}

\pagebreak{}

\section*{Appendices from ``The effect of seasonal strength and abruptness on predator--prey dynamics''}

\global\long\def\thesubsection{Appendix \Alph{subsection}}%
 \setcounter{subsection}{0}

\subsection{Dynamical response to a smooth seasonal forcing on discovery rates}
\label{sec:TS_BifurcDiagEpsilon} 
\global\long\def\thefigure{A\arabic{figure}}
\setcounter{figure}{0}
 
\begin{figure}[H]
\centering
\includegraphics[scale=0.75, trim = {2cm 0 2cm 0}, clip]{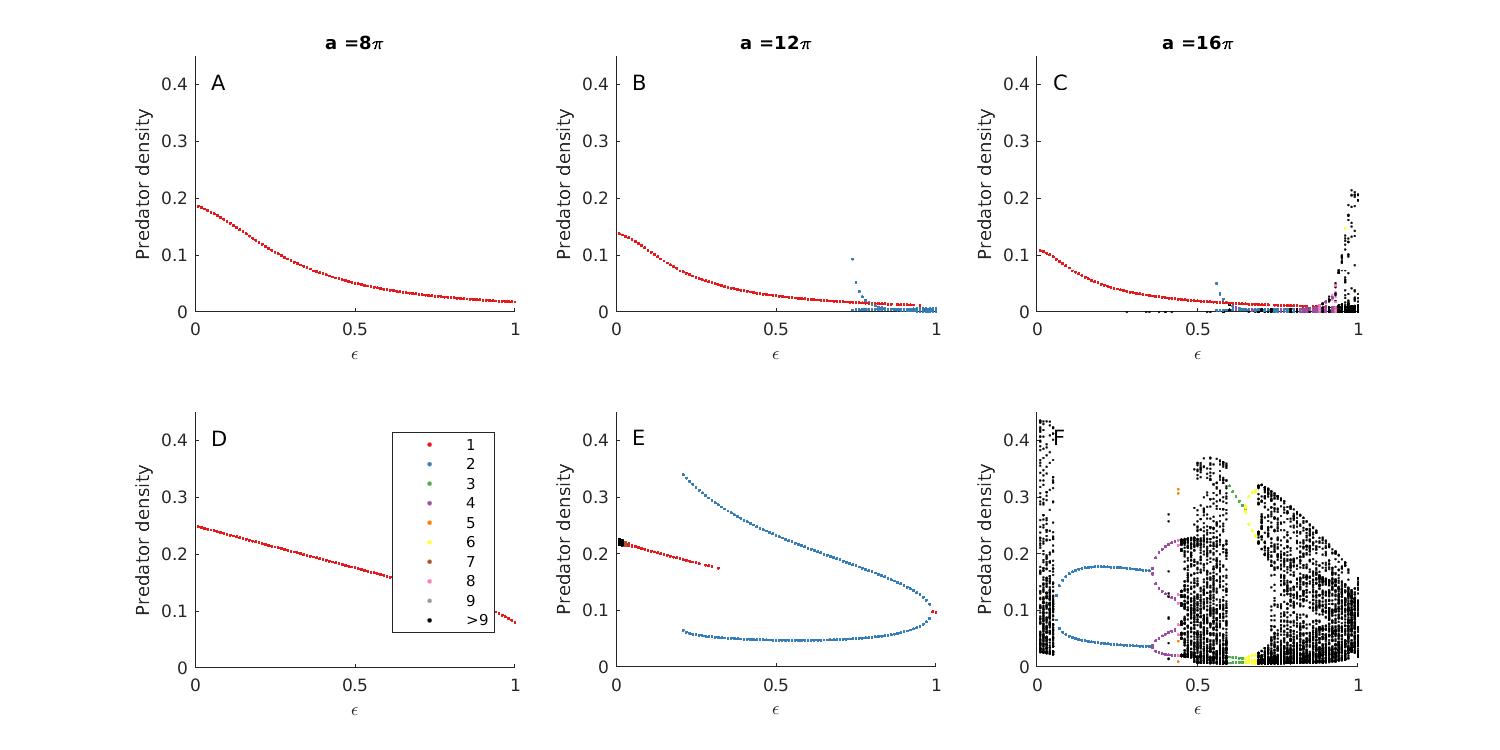}
\caption{Predator density at the beginning of each year (10 years recorded) against the magnitude of the seasonal forcing $\epsilon$, with predation governed by A--B--C) a type I functional response, and D--E--F) a type II functional response. Each plot corresponds to a value of the mean discovery rates $\bar{a}$: $8\pi$ (A and D), $12\pi$ (B and E), and $16\pi$ (C and F). Each point corresponds to a given simulation with a specific set of initial conditions. For each value of $\epsilon$, we ran 10 simulations initiated with different densities ($(x_0,\,y_0) \in ]0,\,1]$). Point colour corresponds to system's periodicity estimated with \citet{taylor_how_2013}'s algorithm, with black points indicating periodicity over 9 years, quasi--periodic or chaotic regimes.}
\label{fig:BifurcDiagOnepsilon_Pred}
\end{figure}

\begin{figure}[H]
\centering
\includegraphics[scale = 0.75]{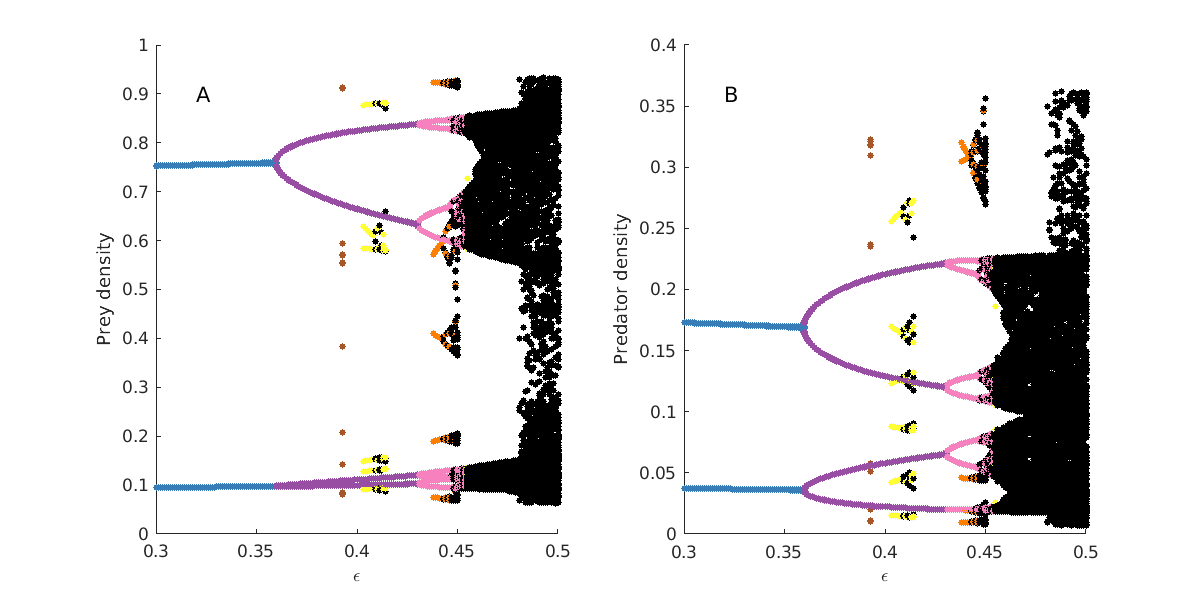}
\caption{Close--up on the bifurcation diagram of Fig. \ref{fig:BifurcDiagOnepsilon_Prey} and \ref{fig:BifurcDiagOnepsilon_Pred}.F for $\epsilon \in [0.3,\,0.5]$. Here, predation is governed by a type II functional response, and the mean discovery rate $\bar{a}$ is $16 \pi$. For each value of $\epsilon$ explored, we ran 20 simulations initiated with different densities.}
\label{fig:ZoomBifurcDiagEpsilon_TypeII_a16pi_E0305}
\end{figure}

\begin{figure}[H]
\centering
\includegraphics[scale = 0.75]{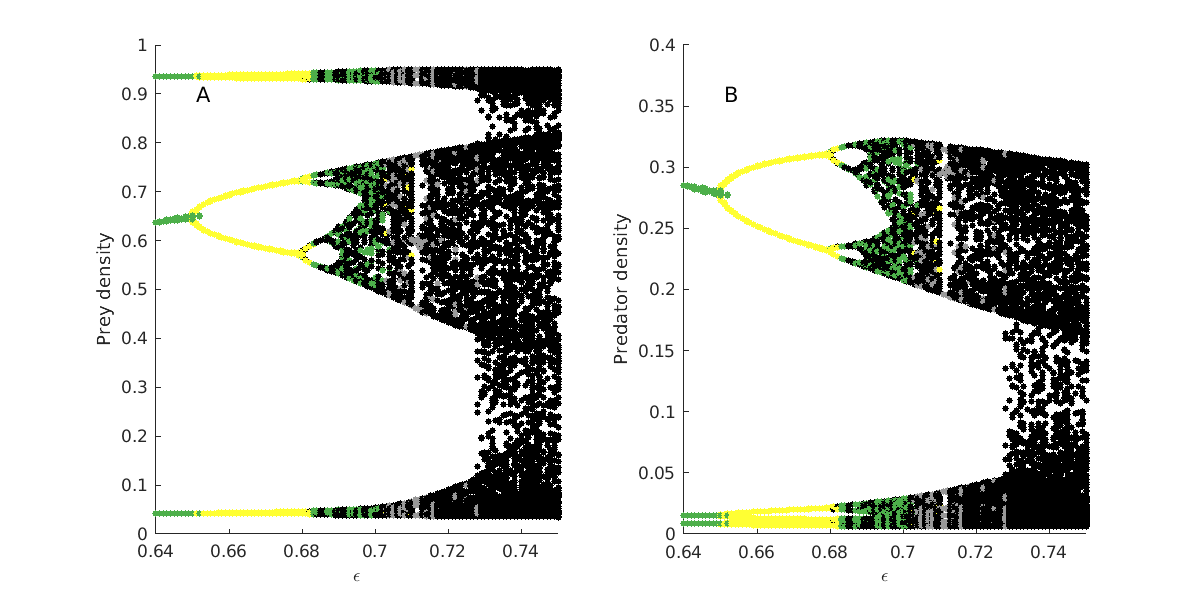}
\caption{Close--up on the bifurcation diagram of Fig. \ref{fig:BifurcDiagOnepsilon_Prey} and \ref{fig:BifurcDiagOnepsilon_Pred}F for $\epsilon \in [0.64,\,0.75]$. Here, predation is governed by a type II functional response, and the mean discovery rate $\bar{a}$ is $16 \pi$. For each value of $\epsilon$ explored, we ran 20 simulations initiated with different densities.}
\label{fig:ZoomBifurcDiagEpsilon_TypeII_a16pi_E065075}
\end{figure}

\begin{figure}[H]
\centering
\includegraphics[scale=0.75]{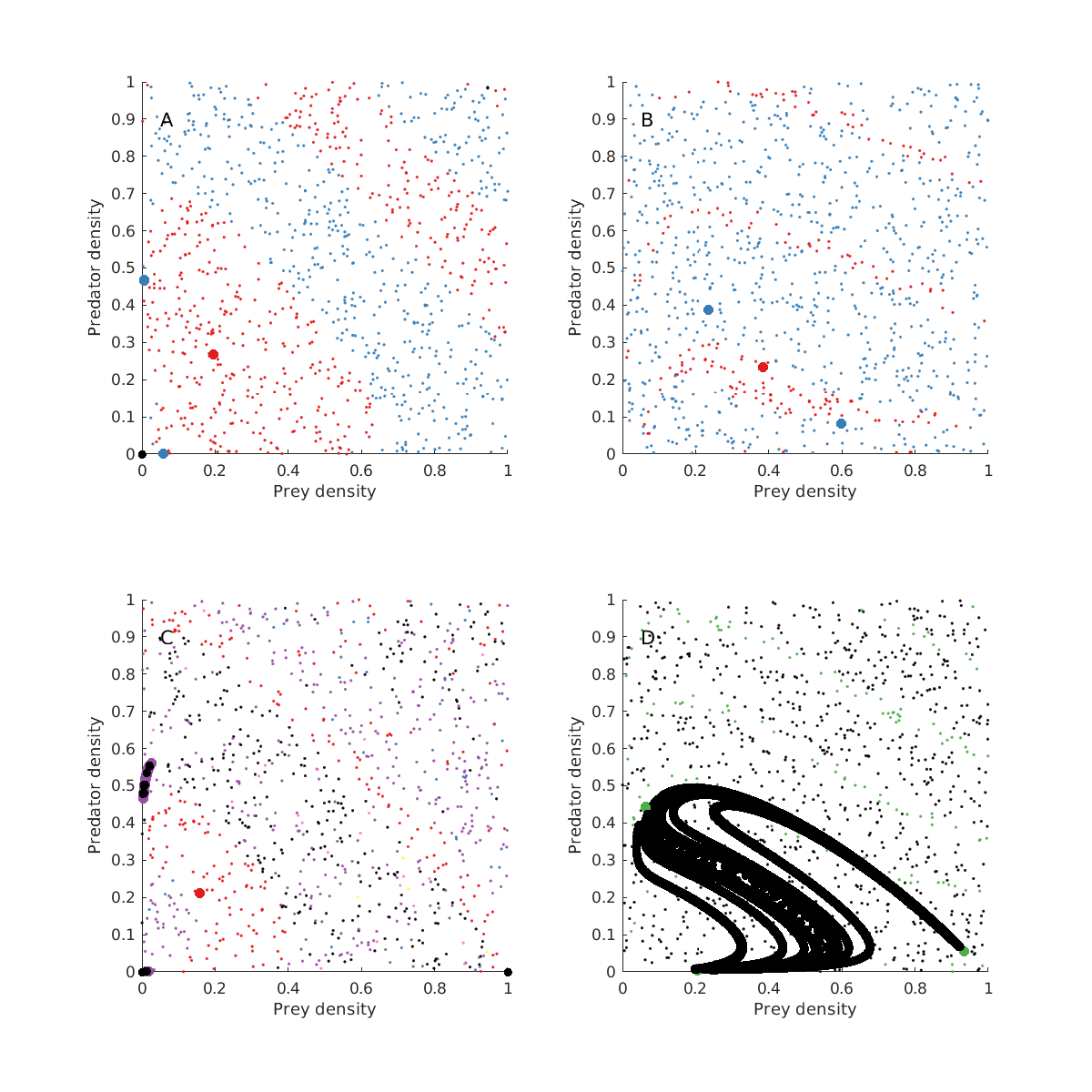}
\caption{Attractors and basins of attraction of the predator--prey models defined by eq. \eqref{eq:system}, \eqref{eq:SeasonalA} and \eqref{eq:period_func} for various strengths of seasonal forcing. Basins are delineated by initial conditions (small dots) of similar colour while attractors are represented by the state of the system at the peak of the predation season ($s(t, \epsilon, \theta)>0$) for the last 100 years (large dots). A) Type I functional response, $\bar{a} = 12 \pi$ and $\epsilon = 0.8$; B) type II functional response, $\bar{a} = 12 \pi$ and $\epsilon = 0.25$; C) type I functional response, $\bar{a} = 16 \pi$, and $\epsilon = 0.82$; D) type II functional response, $\bar{a} = 16 \pi$ and $\epsilon  = 0.59$.}
\label{fig:Bassins}
\end{figure}

\begin{figure}[H]
\centering
\includegraphics{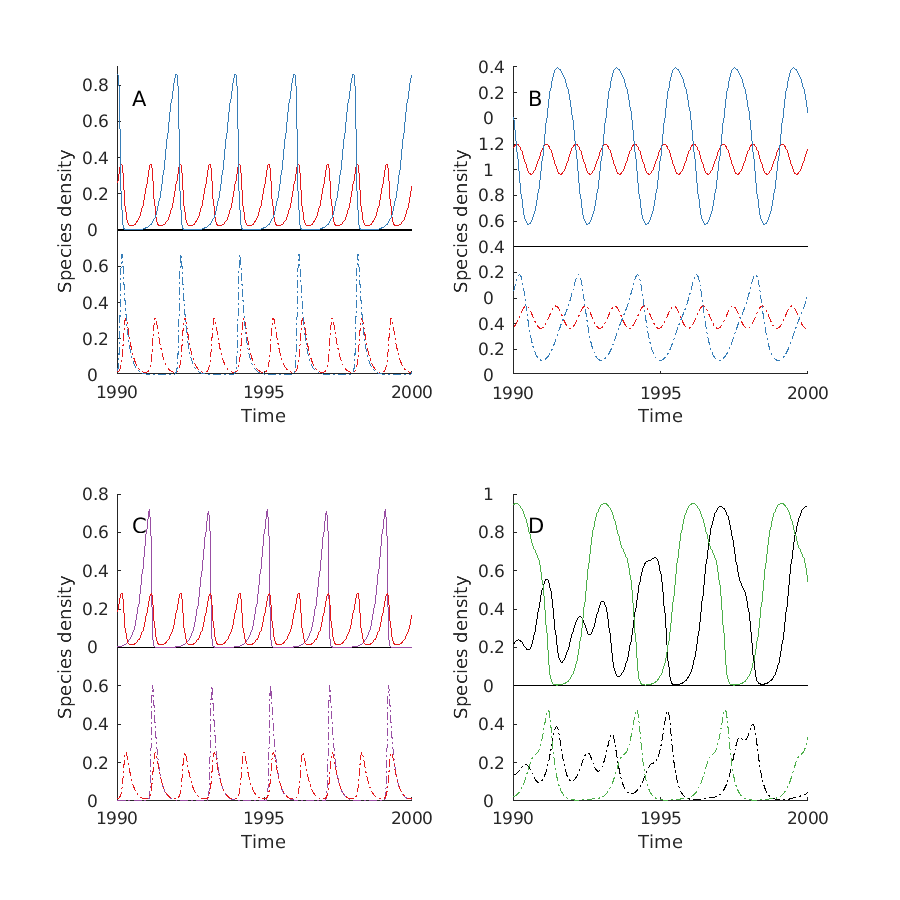}
\caption{Illustrative time series for coexisting attractors with different periodicities.
A) Type I functional response, $\bar{a} = 12 \pi$ and $\epsilon = 0.8$; B) type II functional response, $\bar{a} = 12 \pi$ and $\epsilon = 0.25$; C) type I functional response, $\bar{a} = 16 \pi$, and $\epsilon = 0.82$; D) type II functional response, $\bar{a} = 16 \pi$ and $\epsilon  = 0.59$. Solid lines correspond to prey density and dot--dash lines to predators. Colour code for periodicity is the same as in Fig. \ref{fig:BifurcDiagOnepsilon_Pred}.}
\label{fig:TS_EpsilonOnly}
\end{figure}

\pagebreak

\subsection{Stroboscopic maps of seasonally forced predator–prey dynamics}
\label{sec:StroboscopicMaps} 
\global\long\def\thefigure{B\arabic{figure}}
\setcounter{figure}{0}

\begin{figure}[H]
\centering \includegraphics[scale=0.75, trim = {2cm 0cm 2cm 0}, clip]{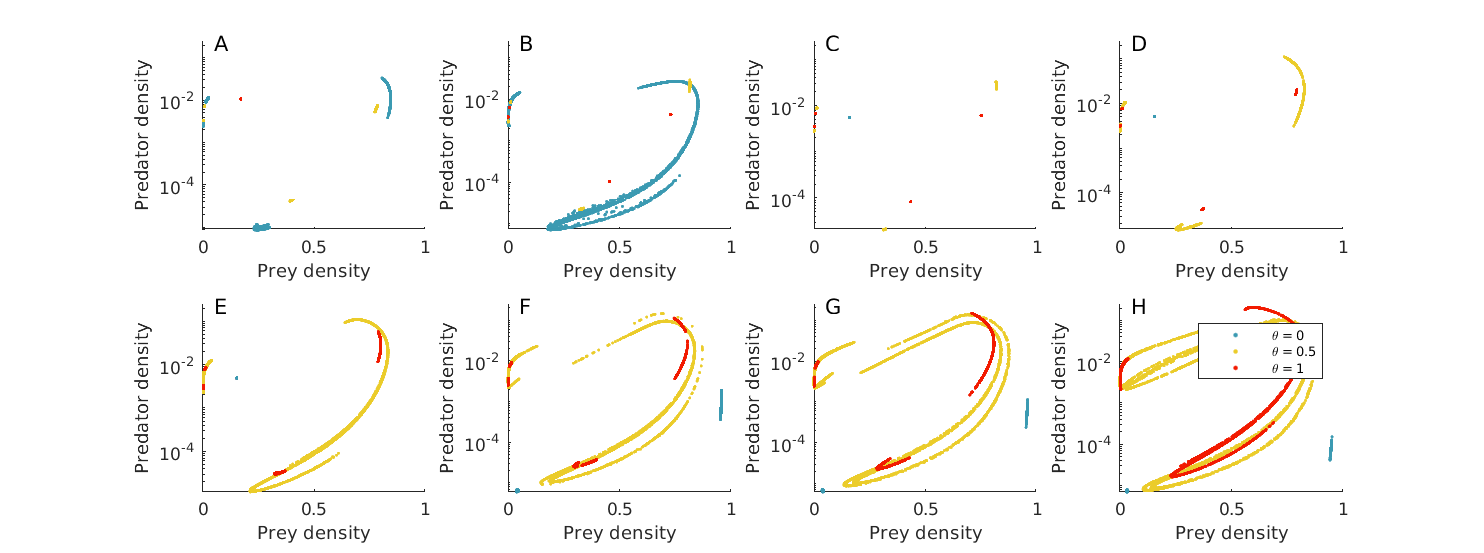}
\caption{Stroboscopic maps of seasonally forced predator--prey dynamics driven by a type I functional response for A) $\epsilon = 0.81$, B) $\epsilon = 0.85$, C) $\epsilon = 0.86$, D) $\epsilon = 0.90$, F) $\epsilon = 0.93$, E) $\epsilon = 0.95$, G) $\epsilon = 0.96$, and H) $\epsilon = 1$.
The community state is displayed for the last 1000 years of simulations, each point corresponding to the beginning of the year, and colours to different shapes of the forcing signal.
Here, simulations are initiated with $(x_0,\,y_0) = (0.3,\,0.3)$ and run for 5000 years, and the mean discovery rate is set at $\bar{a} = 16\pi$.
The y--axis is log--transformed for the sake of clarity.}
\label{fig:SM_ThetaEpsilon_Type1} 
\end{figure}

\begin{figure}[H]
\centering \includegraphics[scale=0.75, trim = {2cm 0cm 2cm 0}, clip]{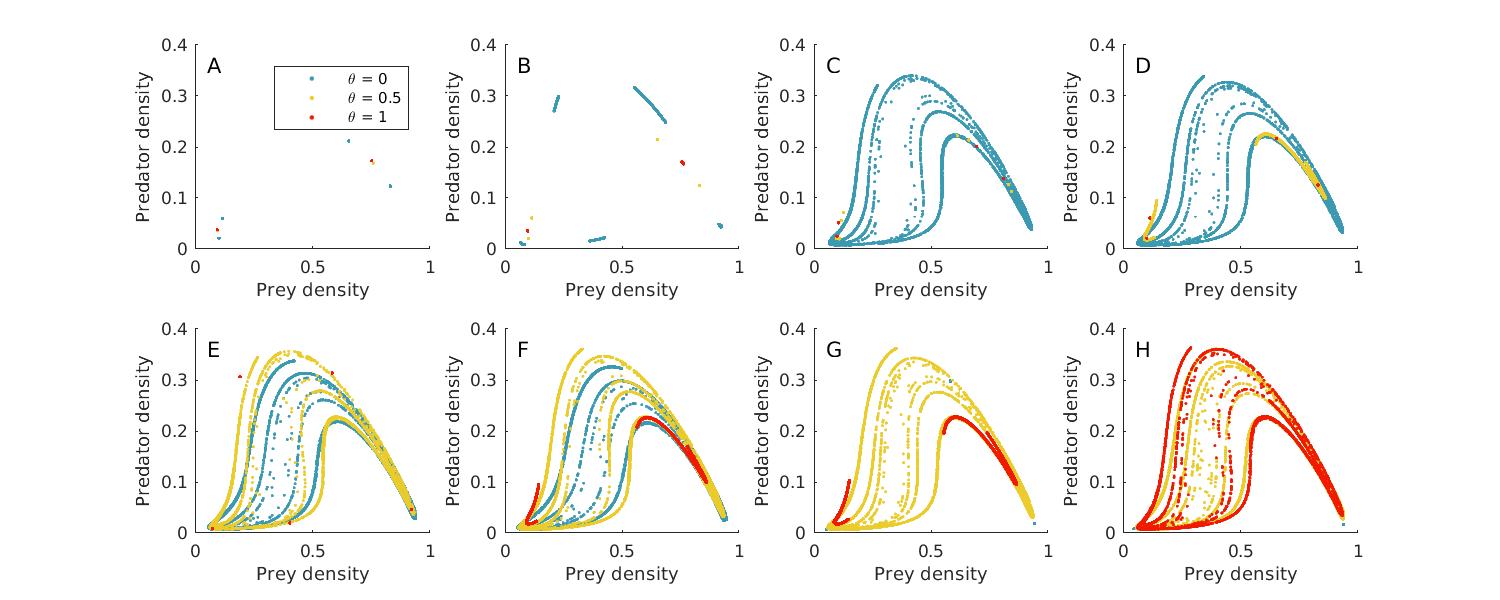}
\caption{Stroboscopic maps of seasonally forced predator--prey dynamics driven by a type II functional response for A) $\epsilon = 0.3$, B) $\epsilon = 0.36$, C) $\epsilon = 0.38$, D) $\epsilon = 0.41$, F) $\epsilon = 0.44$, E) $\epsilon = 0.47$, G) $\epsilon = 0.48$, and H) $\epsilon = 0.5$. The community state is displayed for the last 1000 years of simulations, each point corresponding to the beginning of the year, and colours to different shapes of the forcing signal. Here, simulations are initiated with $(x_0,\,y_0) = (0.3,\,0.3)$ and run for 5000 years, and the mean discovery rate is set at $\bar{a} = 16\pi$.}
\label{fig:SM_ThetaEpsilon_Type2} 
\end{figure}

\begin{figure}[H]
\centering \includegraphics[scale=0.75, trim = {2cm 0 2cm 0}, clip]{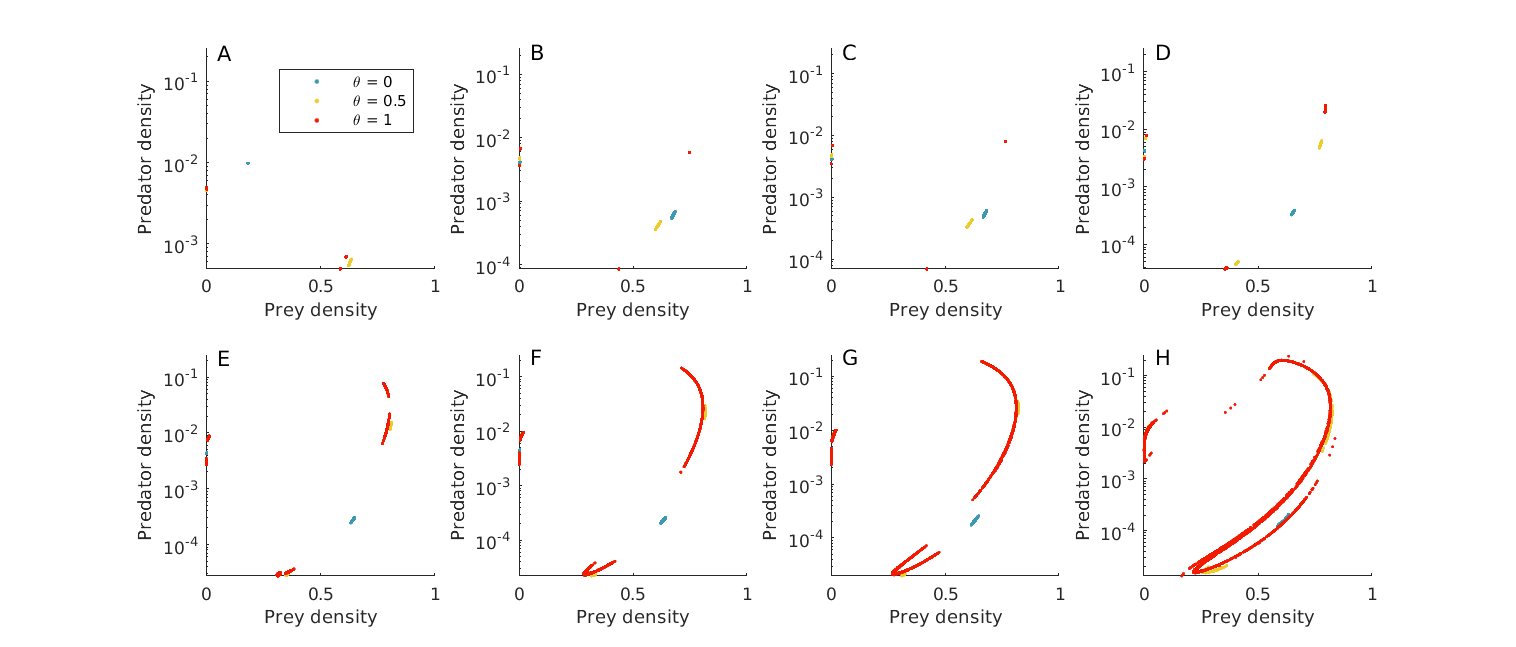}
\caption{Stroboscopic maps of seasonally forced predator--prey dynamics with type I functional response and signal variance controlled: A) $\epsilon = 0.81$, B) $\epsilon = 0.85$, C) $\epsilon = 0.86$, D) $\epsilon = 0.90$, E) $\epsilon = 0.93$, F) $\epsilon = 0.95$, G) $\epsilon = 0.96$, and H) $\epsilon = 1$. The community state is displayed for the last 1000 years of simulations, each point corresponding to the beginning of the year, and colours to different shapes of the forcing signal. Here, simulations are initiated with $(x_0,\,y_0) = (0.3,\,0.3)$ and run for 5000 years, and the mean discovery rate is set at $\bar{a} = 16\pi$.}
\label{fig:SM_CV_ThetaEpsilon_Type1} 
\end{figure}

\begin{figure}[H]
\centering \includegraphics[scale=0.75, trim = {2cm 0 2cm 0}, clip]{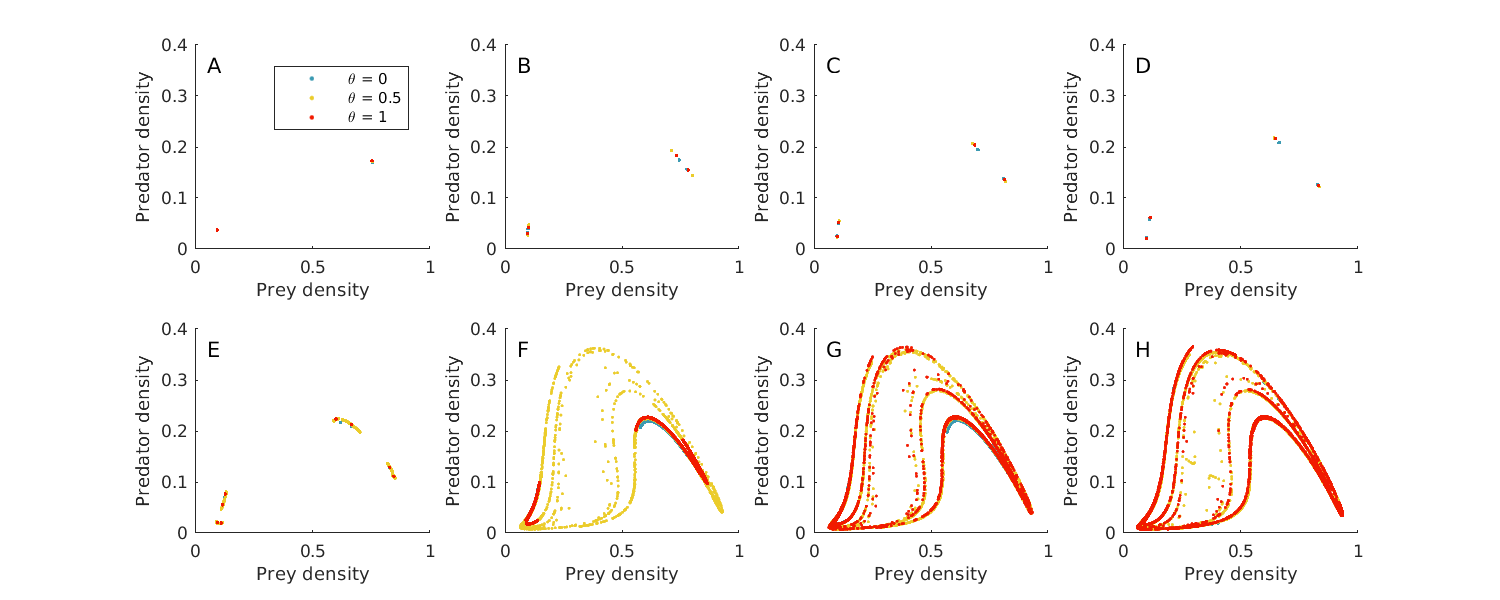}
\caption{Stroboscopic maps of seasonally forced predator--prey dynamics with type II functional response signal variance controlled: A) $\epsilon = 0.3$, B) $\epsilon = 0.36$, C) $\epsilon = 0.38$, D) $\epsilon = 0.41$, E) $\epsilon = 0.44$, F) $\epsilon = 0.47$, G) $\epsilon = 0.48$, and H) $\epsilon = 0.5$. The community state is displayed for the last 1000 years of simulations, each point corresponding to the beginning of the year, and colours to different shapes of the forcing signal. Here, simulations are initiated with $(x_0,\,y_0) = (0.3,\,0.3)$ and run for 5000 years, and the mean discovery rate is set at $\bar{a} = 16\pi$.}
\label{fig:SM_CV_ThetaEpsilon_Type2} 
\end{figure}

\pagebreak{}

\section*{Supplementary information}

\global\long\def\thesubsection{Supplementary information \arabic{subsection}}%
\setcounter{subsection}{0}

\global\long\def\thefigure{S\arabic{figure}}
\setcounter{figure}{0}

\subsection{Unforced predator--prey community dynamics}\label{sec:UnforcedSystem}

\global\long\def\theequation{S1.\arabic{equation}}
\setcounter{equation}{0}

\subsubsection*{Type I functional response}

A functional response of type I assumes that the intake of prey is
proportional to prey density only so it linearly increases with it
($f(x)=ax$, $a$ being the discovery rate). This model allows the
coexistence of both populations,

\begin{equation}
\left\{ \begin{array}{l}
x^{*}=\frac{d}{ac}\\
y^{*}=\frac{r}{a}\left(1-\frac{d}{acK}\right)
\end{array}\right.
\end{equation}

provided that $K>\frac{d}{ac}$, a condition we assume to be satisfied in what follows.

The corresponding Jacobian matrix is

\begin{equation}
J^{*}=\begin{bmatrix}-\frac{rd}{acK} & -\frac{d}{c}\\
cr\left(1-\frac{d}{acK}\right) & 0
\end{bmatrix}
\end{equation}

As the trace is $-\frac{rd}{acK}<0$ and the determinant is $rd\left(1-\frac{d}{acK}\right)>0$, this equilibrium is stable.

The type of equilibrium depends on the sign of $\left(trA^{*}\right)^{2}-4detA^{*}=\frac{rd}{acK}\left(\frac{rd}{acK}-R\left(acK-d\right)\right)$. Two types of stable equilibria are possible (Fig. \ref{fig:EquilibriumState}A, and Fig. \ref{fig:BifurcDiagOna}A):

\begin{itemize}
\item $K<\frac{d}{ac}\left(1+\sqrt{1+\frac{r}{d}}\right)$, the equilibrium is a stable node. Transient dynamics correspond to a monotonic decay to the equilibrium.
\item $K>\frac{d}{ac}\left(1+\sqrt{1+\frac{r}{d}}\right)$, the equilibrium is a stable focus. Transient dynamics correspond to an oscillatory decay to the equilibrium.
\end{itemize}

\subsubsection*{Type II functional response}

A functional response of type II describes an intake of prey which
increases linearly at low prey density and saturates at higher densities:

\begin{equation}
f(x)=\frac{ax}{1+ahx}\label{eq:TypeII_Holling_SI}
\end{equation}

where $h$ is the predator's handling time (in $y$). With $g=\nicefrac{1}{h}$ denoting the maximum per capita predation rate (in $y^{-1}$), and $b=\nicefrac{1}{ah}$ the half saturation constant (in $N.km^{-2}$),
eq. \eqref{eq:TypeII_Holling_SI} becomes

\begin{equation}
f(x)=\frac{gx}{b+x}\label{eq:TypeII}
\end{equation}

which corresponds to the well--known Rosenzweig--MacArthur model \citep{rosenzweig_graphical_1963} in its most widespread form in works on seasonality in predator--prey systems \citep[e.g.,][]{rinaldi_multiple_1993,turchin_empirically_1997,king_rainbow_1999,taylor_seasonal_2013}.

A full--coexistence state corresponds to

\begin{equation}
\left\{ \begin{array}{l}
x^{*}=\frac{d}{ac-dha}\\
y^{*}=\frac{r}{a}\left(1-\frac{x^{*}}{K}\right)\left(1+ahx^{*}\right)
\end{array}\right.\label{eq:EqTypeII}
\end{equation}

Feasibility of the full--coexistence equilibrium requires $\nicefrac{1}{h}>\frac{d}{c}$ and $1-\frac{x^{*}}{K}>0$, i.e. $K>\frac{d}{ac-ahd}$.

To investigate community stability, we adimensionalise the system with $u=ahx$, $v=\frac{ahy}{c}$, and $\tau=\nicefrac{ct}{h}$. The system of ODE \eqref{eq:system} with type II functional response becomes:

\begin{equation}
\left\{ \begin{array}{l}
\frac{du}{d\tau}=Ru\left(1-\frac{u}{C}\right)x-\frac{uv}{u+1}\\
\frac{dv}{d\tau}=\frac{uv}{u+1}-Qv
\end{array}\right.
\end{equation}

where $R=\frac{rh}{c}$, $C=ahK$, and $Q=\nicefrac{cd}{h}$.

The full equilibrium \eqref{eq:EqTypeII} translates into

\begin{equation}
\left\{ \begin{array}{l}
u^{*}=\frac{Q}{1-Q}\\
v^{*}=\frac{RQ}{(1-Q)^{2}}\frac{1}{C}\left(C-QC-Q\right)
\end{array}\right.\label{eq:EqTypeII_Adim}
\end{equation}

and the Jacobian matrix at equilibrium is written:

\begin{equation}
J^{*}=\begin{bmatrix}\frac{RQ}{C(1-Q)^{2}}\left(C-QC-Q-1\right) & -Q\\
\frac{R}{C}\left(C-QC-Q\right) & 0
\end{bmatrix}
\end{equation}

\begin{figure}[H]
\centering \includegraphics[scale=0.85, trim = {1cm 0 0 0}, clip]{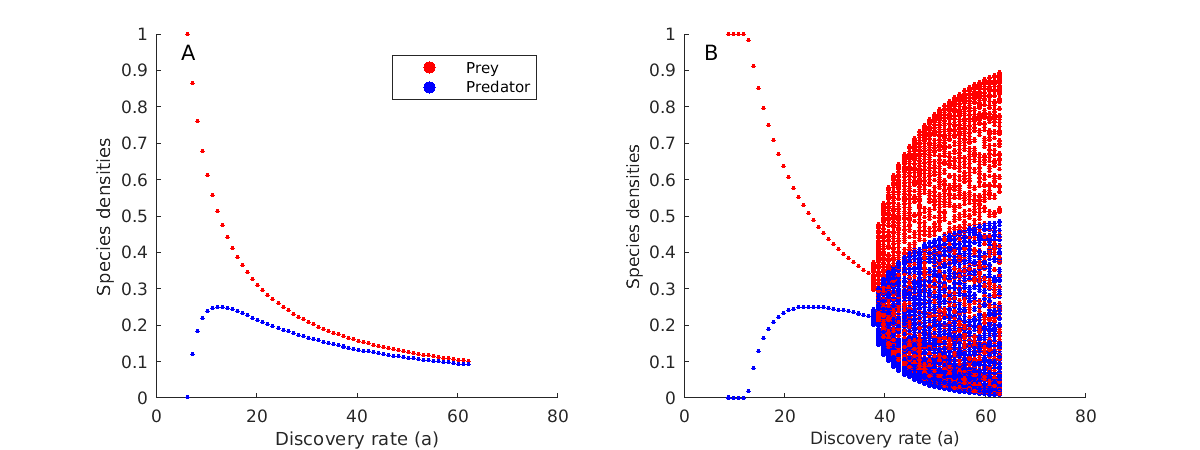}
\caption{Densities of the prey and the predator at the beginning of each year
(20 years recorded) against the discovery rate $a$, with predation
governed by A) a type I functional response, and B) a type II functional
response. For each discovery rate value, we simulate 25 predator--prey
communities which differ only by the initial conditions ($\{x_{0},\,y_{0}\} \in ]0,\,1]$).}
\label{fig:BifurcDiagOna}
\end{figure}

Hence, stability of full--equilibrium requires:

\begin{equation}
\left\{ \begin{array}{l}
trA^{*}=\frac{RQ}{C(1-Q)^{2}}\left(C-QC-Q-1\right)<0\\
detA^{*}=\frac{RQ}{C}\left(C-QC-Q\right)>0
\end{array}\right.
\end{equation}

which equates to $\frac{Q}{1-Q}<C<\frac{Q+1}{1-Q}$, i.e. $\frac{d}{ac-ahd}<K<\frac{hd+c}{ah(c-hd)}$ (Fig. \ref{fig:EquilibriumState}B). Under these conditions, the community converges to a stable equilibrium, either by monotonic or oscillatory decay. When $K\geqslant\frac{hd+c}{ah(c-hd)}$, predator--prey dynamics are characterised by limit cycles (Fig. \ref{fig:BifurcDiagOna}B).

\end{document}